%% file: A2146_HlavacekLarrondo.tex
\documentclass[useAMS,usenatbib]{mn2e}\usepackage{natbib}

\usepackage[T1]{fontenc}
\usepackage{ae,aecompl}
\usepackage[latin1]{inputenc}
\usepackage{graphicx}
\usepackage{float}
\usepackage{color}
\usepackage{footnote}
\usepackage{amssymb}
\usepackage[figuresright]{rotating}
\usepackage{amsmath}
\usepackage{url}
\usepackage{rotating}
\usepackage{color}
\usepackage{soul}

\input{defn} 
\def\ltsim{\mathrel{\hbox{\rlap{\hbox{\lower4pt\hbox{$\sim$}}}\hbox{$<$}}}}
\def\gtsim{\mathrel{\hbox{\rlap{\hbox{\lower4pt\hbox{$\sim$}}}\hbox{$>$}}}}

\title[VLA Observations of Abell 2146]{Mystery solved: discovery of extended radio emission in the merging galaxy cluster Abell 2146}

\author[J. Hlavacek-Larrondo et
al.]{J. Hlavacek-Larrondo$^{1}$\thanks{E-mail:
    juliehl@astro.umontreal.ca}, M.-L. Gendron-Marsolais$^{1}$, D. Fecteau-Beaucage$^{1}$, R. J. \newauthor{van Weeren$^{2}$, H. R. Russell$^{3}$, A. Edge$^{4}$, M. Olamaie$^{5,6}$, C. Rumsey$^{5}$, L. King$^{7}$, } \newauthor{A. C. Fabian$^{3}$, B. McNamara$^{8,9}$, M. Hogan$^{8}$, M. Mezcua$^{1}$, G. Taylor$^{10}$}\\
$^{1}$D\'{e}partement de Physique, Universit\'{e} de Montr\'{e}al, Montr\'{e}al, QC H3C 3J7, Canada\\
$^{2}$Harvard-Smithsonian Center for Astrophysics, 60 Garden Street, Cambridge, MA 02138, USA\\
$^{3}$Institute of Astronomy, University of Cambridge, Madingley Road, Cambridge, CB3 0HA, UK\\
$^{4}$Centre for Extragalactic Astronomy, Department of Physics, Durham University, Durham, DH1 3LE, UK\\
$^{5}$Cavendish Laboratory, Battcock Centre for Experimental Astrophysics, JJ Thomson Avenue, Cambridge, CB3 0HE, UK\\
$^{6}$Imperial Centre for Inference and Cosmology(ICIC), Imperial College, Prince Consort Road, London SW7 2AZ, U.K.\\
$^{7}$Department of Physics, University of Texas at Dallas, 800 West Campbell Road Richardson, Texas 75080-3021, USA \\
$^{8}$Department of Physics and Astronomy, University of Waterloo, Waterloo, ON, N2L 3G1, Canada\\
$^{9}$Perimeter Institute for Theoretical Physics, Waterloo, ON, N2L 2Y5, Canada\\
$^{10}$Department of Physics and Astronomy, University of New Mexico, Albuquerque, NM 87131, USA\\}

\date{Accepted XXX. Received YYY; in original form ZZZ}
\pubyear{2017}
\begin{document}

\label{firstpage}
\pagerange{\pageref{firstpage}--\pageref{lastpage}}
\maketitle

\begin{abstract}
\noindent Abell~2146 ($z=0.232$) is a massive galaxy cluster currently undergoing a spectacular merger in the plane of the sky with a bullet-like morphology. It was the first system in which both the bow and upstream shock fronts were detected at X-ray wavelengths (Mach~$\sim2$), yet deep GMRT 325~MHz observations failed to detect extended radio emission associated with the cluster as is typically seen in such systems. We present new, multi-configuration $1-2$~GHz Karl G. Jansky Very Large Array (VLA) observations of Abell~2146 totalling 16 hours of observations. These data reveal for the first time the presence of an extended ($\approx850$ kpc), faint radio structure associated with Abell~2146. The structure appears to harbour multiple components, one associated with the upstream shock which we classify as a radio relic and one associated with the subcluster core which is consisted as being a radio halo bounded by the bow shock. The newly detected structures have some of the lowest radio powers detected thus far in any cluster ($P_{\rm 1.4 GHz,~halo}=2.4\pm0.2\times10^{23}$~W~Hz$^{-1}$ and $P_{\rm 1.4 GHz,~relic}=2.2\pm0.2\times10^{23}$~W~Hz$^{-1}$). The flux measurement of the halo, as well as its morphology, also suggest that the halo was recently created ($\approx0.3$ Gyr after core passage), consistent with the dynamical state of the cluster. These observations demonstrate the capacity of the upgraded VLA to detect extremely faint and extended radio structures. Based on these observations, we predict that many more radio relics and radio halos in merging clusters should be detected by future radio facilities such as the Square Kilometre Array (SKA). 
\end{abstract}


\section{Introduction}

Clusters of galaxies originate from the hierarchical mergers of
subclusters and groups. As the structures collide, they can reach
velocities of a few thousand km s$^{-1}$ and kinetic energies in
excess of
10$^{64}$ erg. They are ideal laboratories for studying various physical
processes, including key evidence of dark matter \citep{Clo2004604,Mar2004606,Bra2006652,Clo2006648}
and non-thermal processes such as the generation of shocks and turbulence. However, in order
to study these phenomena in detail, an inclination along the plane
of the sky is needed, and having such a favourable angle is
rare. Until recently, only a handful of such clusters were known,
including 1E 0657-55.8 \citep[also known as the Bullet cluster; e.g.][]{Mar2002567}, CIZA J2242.8+5301 \citep[e.g.][]{vanW2011418}, Abell 520
\citep[e.g.][]{Mar2005627} and Abell 754 \citep[e.g.][]{Mac2011728}.

\begin{figure*}
\centering
\begin{minipage}[c]{0.99\linewidth}
\centering \includegraphics[width=\linewidth]{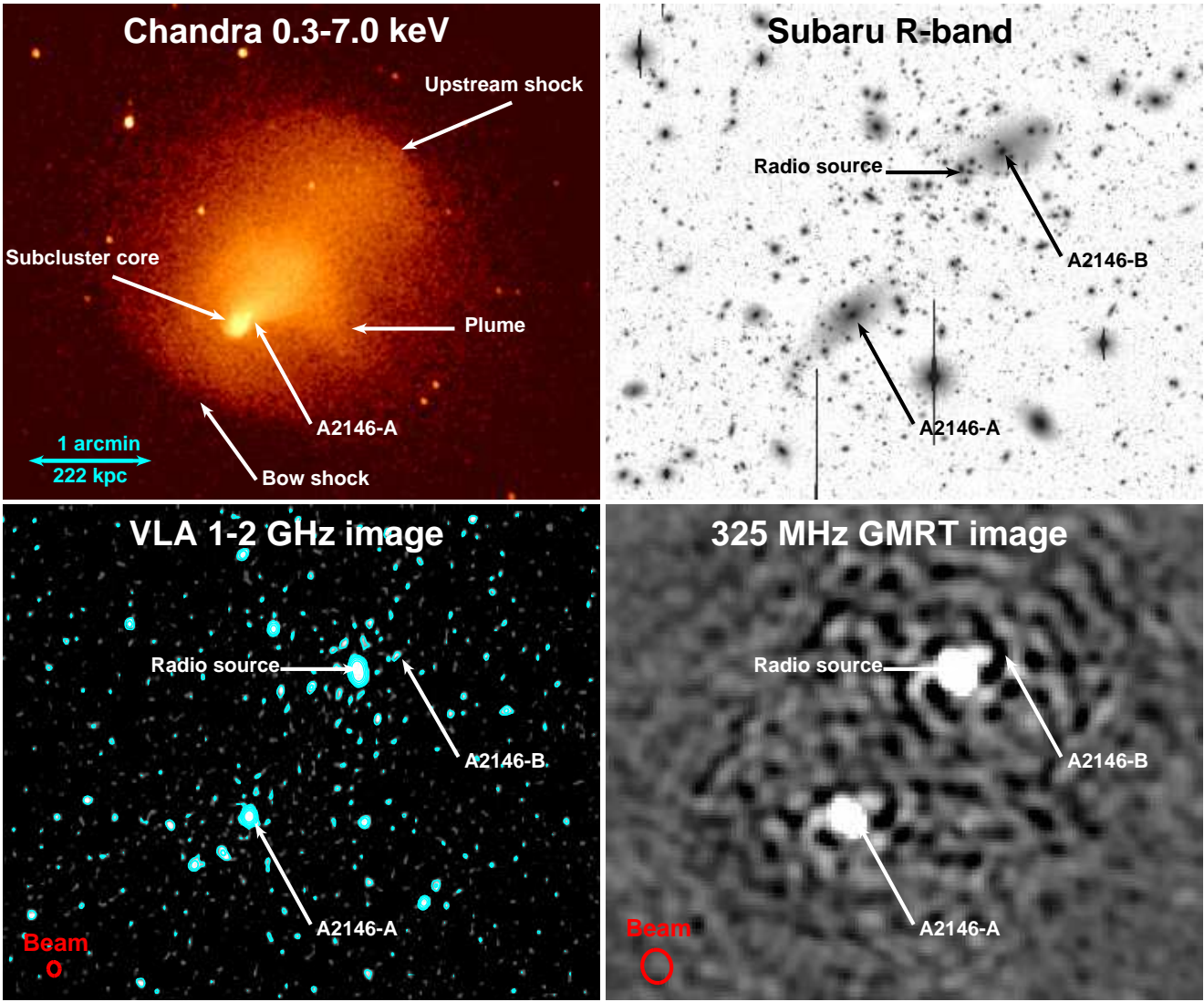}
\end{minipage}
\caption[]{All images are on the same scale. \textbf{Top-left:} \emph{Chandra} image
  in the $0.3-7.0$ keV band (418 ks) from \citet{Rus2012422}, smoothed with a gaussian function of 3 pixels. We highlight several features, including the two shock fronts and BCG A2146-A. \textbf{Top-right:} Subaru Suprime-Cam image in R band from \citet{Whi2015453}. \textbf{Bottom-left:} High-resolution $1-2$ GHz VLA image, combining all datasets and configurations (see column 5 of Table \ref{tab1} for details). The noise level in the vicinity of the cluster is $\sigma_{\rm rms}=11\mu$Jy beam$^{-1}$, with a beam size of $3.9"\times3.2"$. Contours levels are drawn at $[1,3,9,27,...]\times3\sigma_{\rm rms}$ (8 levels in total). We highlight the location of the two BCGs (A2146-A and A2146-B), as well as the northern radio source which coincides with a member galaxy of the cluster. \textbf{Bottom-right:} GMRT 325 MHz radio image from \citet{Rus2011417}. Note the remaining artefacts surrounding the A2146-A and the northern radio source. The beams for the radio images are shown in red in the lower-left corners.}
\label{fig1}
\end{figure*}

We have discovered another such cluster, Abell 2146 \citep[$z =
0.232$, hereafter A2146;][]{Rus2010406,Rus2012423}. It is undergoing a spectacular merger in the plane of the sky
with a similar structure to the Bullet cluster. The \emph{Chandra} image,
shown here in the top-left panel of Fig. \ref{fig1}, reveals the remnants of a cool core (i.e. the bullet) being stripped of its gas and
forming a tail of material leading back to a separate concentration
of gas 400 kpc away. More importantly, the X-ray observations
show that there are unambiguous density and temperature jumps associated with two X-ray shock fronts (Mach $\sim2$): one preceding the cool core
(the bow shock), and another propagating outwards from the primary
cluster (the upstream shock). Clear observations of shock fronts are extremely rare
since they require the merger axis to be close to the
plane of the sky so that projection effects do not smear out the
surface brightness edges \citep[see also][]{Mar2002567,Mar2005627,Mac2011728,Owe2011728}.

A2146 also hosts a brightest cluster galaxy (BCG; labelled A2146-A in Fig. \ref{fig1}) which is
located 10 arcsec ($\approx40$ kpc) behind the X-ray surface brightness peak, placing it
behind the bow shock front. This is unexpected because cluster
galaxies should be collisionless, as opposed to the hot intracluster
gas which is slowed down by friction \citep[e.g.][]{Mar2001563}. In a merger, the galaxies are
therefore expected to preceed the hot X-ray gas. This is not the case
in A2146, and the location of the BCG is unlikely due to projection
effects since the observations of shock fronts require the merger axis
to be close to the plane of the sky \citep[see
also][]{Whi2015453}. Instead, Canning et al. (2012) and White et al. (2015) \nocite{Can2012420,Whi2015453} argue that it
may be due to another galaxy perturbing the system, or that the merger
is slightly off-axis. 

Using 5.7 hours of observations from the Giant MetreWave Telescope (GMRT) at 325 MHz, \citet{Rus2011417} found no evidence of extended radio emission in A2146 down to a rms of 250 $\mu$Jy beam$^{-1}$. The only two radio sources detected
coincided with A2146-A and a second cluster galaxy to the north (labelled as $radio~source$ in Fig. \ref{fig1}, see bottom-right panel). Diffuse, cluster-wide radio emission is
often seen in merging systems, classified as either radio halos or
radio relics depending on their morphology and polarisation
properties \citep[see for a review][]{Bru201423}. Radio
relics are elongated, polarized structures that often trace
shock fronts where particles are being re-accelerated \citep[e.g.][]{Enb1998332}. Radio
halos on the other hand are thought to originate naturally from
shocks and turbulence generated during a merger. These amplify the
magnetic fields and can re-accelerate existing populations of particles. This in return provides a
source of non-thermal emission that can be seen at radio
wavelengths \citep[e.g.][]{Bru2001320,Pet2001557}. Previous observations of cluster mergers have suggested a
strong correlation between the radio power of halos and the X-ray
luminosity of the cluster, known as the $L_{\rm X}-P$ correlation \citep[e.g.][]{Lia2000544,Cas2006369}. However, observations have shown that there
is a bimodal population in this correlation. The cluster halos with weaker radio power for a
given X-ray luminosity are mostly thought to be the aged counterparts of
those that follow the correlation \citep[e.g.][]{Don2013429}. The non
detection of a radio halo for A2146 by \citet{Rus2011417} might therefore indicate that this cluster fits
well into this aged population. However, this is inconsistent with the
X-ray observations since the estimated age from the shock fronts puts
the cluster at an evolutionary stage of $0.2-0.3$ Gyr after core
passage \citep[see][]{Rus2010406,Rus2012423,Whi2015453}, similar to the Bullet cluster, yet the Bullet cluster clearly
sits along the correlation. 

Radio halos can also be produced by inelastic hadronic collisions
between cosmic-ray protons and thermal protons \citep[see e.g.][]{Den1980239}. In this scenario, the radio luminosities of radio halos should be around 10 times less than expected from re-acceleration models, providing a possible explanation for the weaker radio power sources in the $L_{\rm X}-P$ correlation \citep[see also][]{Bro2011740}. These models also predict that magnetic fields should be different in clusters with or without halos and that radio halo hosting clusters should be gamma-ray luminous \citep[e.g.][]{Jel2011728}. Both of these predictions have not been observed \citep[e.g.][]{Bon2011530,Ack2014787}.

Prior to A2146, all merging clusters of galaxies
with X-ray detected shock fronts had large diffuse radio emission
associated with them. The non-detection in A2146 was therefore puzzling
and deeper radio observations were needed to confirm the lack of
extended radio emission. In this paper, we present new, multi-configuration Karl G. Jansky Very Large Array (VLA)
observations of A2146 at $1-2$ GHz. In Section 2, we describe the
observations, and then we present our main result in Section 3,
the detection of extended radio emission in A2146. In
Section 4, we discuss the implications of these results. Finally, in
Section 5 we state our conclusions. Throughout this paper, we adopt
$H_\mathrm{0}=70\kmpspMpc$ with $\Omm=0.3$, $\OmL=0.7$, resulting in a
scaling of 3.698 kpc per arcsec at the redshift of the cluster ($z=0.232$). All errors are $1\sigma$ unless otherwise noted.

\begin{table*}
\caption[]{Radio observations of the merging galaxy cluster A2146.}
\hspace{-0.31in}
\resizebox{18cm}{!} {
\begin{tabular}{lccccccc}
\hline
& VLA B-array & VLA C-array & VLA D-array & Combined VLA & GMRT \\
\hline
Observing dates & 2012 June 9 & 2012 March 16  & 2013 January 27 & ... & 2010 August 22  \\
&   &   2012 April 22 & & & \\
On source time  & 2.4 hrs & 7.4 hrs & 1.3 hrs & 11.1 hrs & 5.7 hrs  \\
Correlations  & Full Stokes & Full Stokes & Full Stokes & Full Stokes & RR and LL\\
Frequency & $1-2$ GHz & $1-2$ GHz & $1-2$ GHz & $1-2$ GHz & $309-357$ MHz \\
Beam size & $3.7"\times2.9"$& $9.9"\times7.7"$ & $47.1"\times30.0"$ & $3.9"\times3.2"$ & $9.3"\times8.1"$ \\
Beam PA & $-8.1^o$ & 48.5$^o$ & $28.0^o$ & $-3.7^o$ &  $10.7^o$ \\
rms & 11 $\mu$Jy beam$^{-1}$& 11 $\mu$Jy beam$^{-1}$& 64 $\mu$Jy beam$^{-1}$ & 11 $\mu$Jy beam$^{-1}$ & 250 $\mu$Jy beam$^{-1}$\\
\hline
\end{tabular}}
\label{tab1}
\end{table*}

\section{Observations and data reduction}

\subsection{VLA observations}

Four observations of A2146 were carried out in L-band on the
VLA between 16 March 2012 and 27 January 2013 (PI Hlavacek-Larrondo). Two of these were in
C-array, one in B-array
and one in D-array. The L-band covers a frequency range from 1.0 to
2.0 GHz, with two 512 MHz IF pairs,
each comprising 8 subbands of 64 MHz. We used the default
integration time for the L-band (3 seconds for B-array and
5 seconds for C-array and D-array). The data were recorded in 16 spectral windows, each subdivided into
64 channels. Full stokes correlations were recorded for a total
acquisition time of 16 hours. Details of each dataset can be found in Table \ref{tab1}.

Data reduction was achieved
using CASA (v.4.6.0), following the procedure described in the L-band
tutorial found on-line on the NRAO
website\footnote{https://casaguides.nrao.edu/index.php/Karl$\_$G.$\_$Jansky$\_$VLA$\_$
  Tutorials}. Briefly,
malfunctioning antennas as
well as radio frequency interference (RFI) were flagged based
on the observation logs and visual inspection of the
datasets. The data were then Hanning smoothed. After a first bandpass
correction,  automative flagging routines (TFCrop, RFlag) were
applied. The data were calibrated using the normal calibration
tasks (\textsc{setjy}, delay and bandpass calibrations, gain calibration and
flux scaling the gain solutions). The calibrations were then applied
and the corrected data were splitted. For all
observations, 3C286 was used as the primary calibrator, while J1634+6245 was used as a phase calibrator.
\begin{figure*}
\centering
\begin{minipage}[c]{0.9\linewidth}
\centering \includegraphics[width=\linewidth]{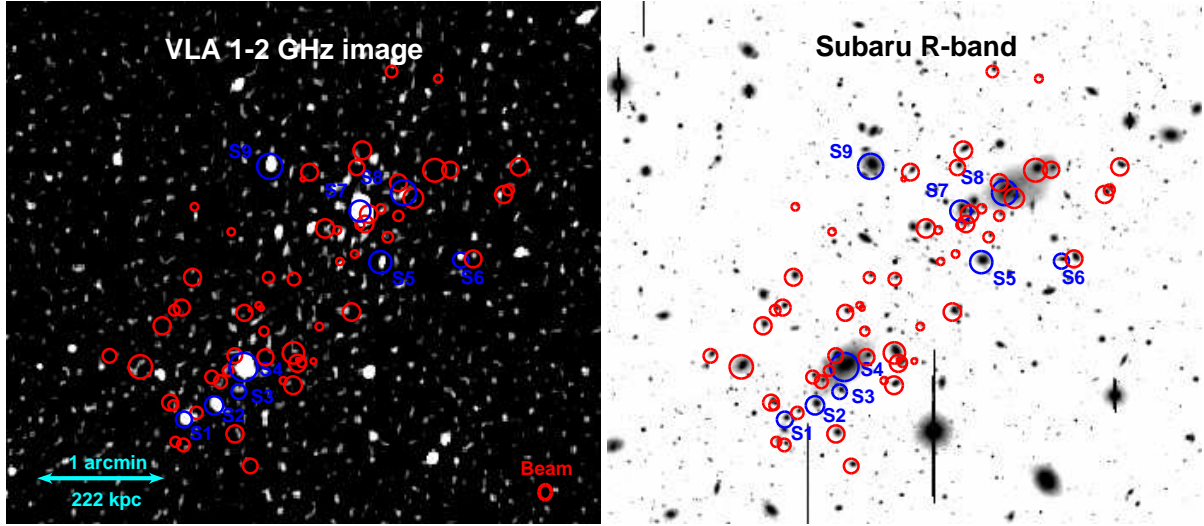}
\end{minipage}
\caption[]{\textbf{Left:} High-resolution $1-2$ GHz VLA image (same as in Fig. \ref{fig1}). In blue and red, we highlight the 68 spectroscopically confirmed cluster members found by \citet{Whi2015453} and Coleman et al. (submitted), The blue regions indicate those that coincide with a radio source detected above a $3\sigma_{\rm rms}$ level (labelled S1 to S9 in the right panel), where $\sigma_{\rm rms}=11\mu$Jy beam$^{-1}$. \textbf{Right:} Subaru Suprime-Cam R-band image (same as in Fig. \ref{fig1}).}
\label{fig2}
\end{figure*}

For each dataset, three rounds of phase-only self-calibration were
performed in order to improve calibration
accuracy. Due to the high number of radio point sources, the software
PyBDSM \citep[v.1.8.6][]{Moh2015} was used in order to create island-type masks, which
were inspected and included in the \textsc{clean} process. For each of the imaging rounds during the self-calibration, we used \citet{Bri1995} weighting, multi-frequency synthesis (MS-MFS) with \textsc{nterms=2} \citep{Rau2011532}, and
480 W-projection planes \citep{Cor2005347}. We also used the
\textsc{robust} factor set to $-0.5$ for these images. The resulting properties, for each configuration, are shown in Table \ref{tab1}. Note that the noise level (rms) shown for all datasets represents the rms located near the cluster. When considering a region located far from the cluster, the rms was typically a factor 2 smaller and reached the thermal noise level. We also note that the largest angular scales detectable are $\approx8$ arcminutes for B-array, $\approx70$ arcminutes for C-array and $\approx70$ arcminutes for D-array.

The final step consisted in merging the datasets together, and producing a final combined (B-array, C-array, D-array) image of the cluster. First, we combined the two measurements sets in C-array to produce a final image in that configuration. Here, the corrected column of each self-calibrated measurement set was split and then merged by running a \textsc{clean} process. We then applied a long 10 minute phase-only calibration and finally produced a combined image with \textsc{clean}. The extra calibration step allowed us to correct for mis-alignments between the datasets. A similar procedure was then applied to datasets from all 3
configurations (B-array, C-array and D-array), and a final, high-resolution image was produced (shown in the lower-left panel of Fig. \ref{fig1}). When producing this final image, we maximise the resolution (oversampling the beam by a factor of at least 3) and we use \textsc{Briggs} weighting, MS-MFS with \textsc{nterms=2}, 480 W-projection planes and a \textsc{robust} factor of $-0.5$. The resulting rms near the cluster of this combined image is 11 $\mu$Jy beam$^{-1}$, but goes down to $5-6\mu$Jy beam$^{-1}$ when measured far from the source. Note that the final images were corrected for
the primary beam. 

\begin{figure*}
\centering
\begin{minipage}[c]{0.9\linewidth}
\centering \includegraphics[width=\linewidth]{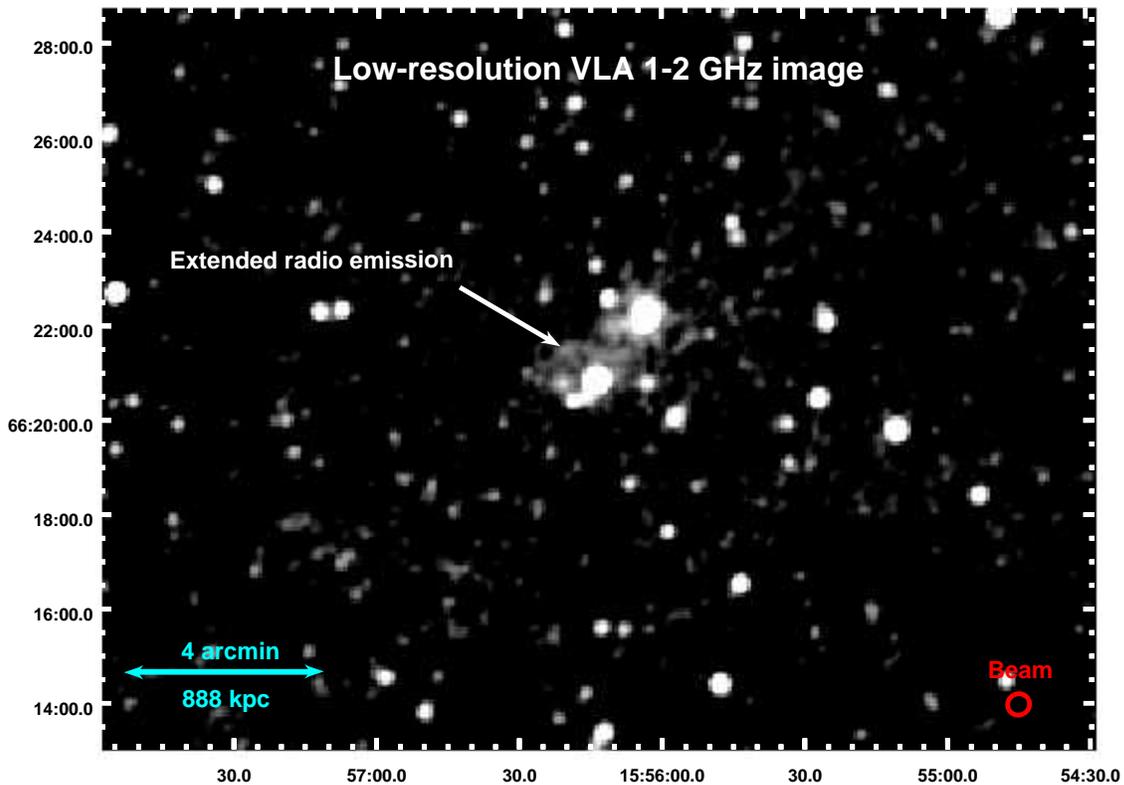}
\end{minipage}
\caption[]{Large-scale low-resolution $1-2$ GHz image obtained with the
  VLA observations, combining all datasets and revealing the presence of extended radio emission centered on A2146. The noise level is $\sigma_{\rm rms}=12\mu$Jy beam$^{-1}$.}
\label{fig3_large}
\end{figure*}

\subsection{GMRT observations}

GMRT continuum observations were obtained at 325 MHz for A2146
on 22 August 2010 (5.7 hours on source), and are presented in detail in \citet{Rus2011417}. These observations were taken
with the software backend correlator with a total bandwidth of 32 MHz
divided between 512 channels. The data were reduced using \textsc{AIPS} and the
deepest image constructed gave a rms noise far from any sources of
92$\mu$Jy beam$^{-1}$, with beam size of $9.3''\times8.1''$. However, the rms near the
target goes up to 250 $\mu$Jy beam$^{-1}$ due to remaining artefacts in the
image.  The bottom-right panel of Fig. \ref{fig1} shows the resulting image at 325
MHz. The observations are summarized in Table \ref{tab1}. Additional details regarding the data reduction can be found in
\citet{Rus2011417}.

\subsection{\emph{Chandra} X-ray observations}

Throughout this paper, we also make use of \emph{Chandra} observations of
A2146. The details of these observations can be found in
\citet{Rus2012422}. Briefly, A2146 was observed with the ACIS-I
detector for a total of 377 ks (ObsIDs 10888, 12245, 13120, 12246, 13138, 12247, 13023, 13020 and 13021) and the ACIS-S detector for 39 ks (ObsID 10464). The data were processed, cleaned and calibrated using {\sc ciao} ({\sc
  ciaov4.3}, {\sc caldb 4.4.0}), starting from the level 1 event
file. Charge transfer inefficiency and time-dependent gain corrections
were applied and flares were removed using the {\sc {lc$\_$clean}}
script. Datasets were reprojected to match the ObsID 12245 dataset,
combined and then exposure-corrected to produce the image shown in
the top-left panel of Fig. \ref{fig1}. The net exposure time is 418 ks.

\section{Results}

\begin{figure*}
\centering
\begin{minipage}[c]{0.9\linewidth}
\centering \includegraphics[width=\linewidth]{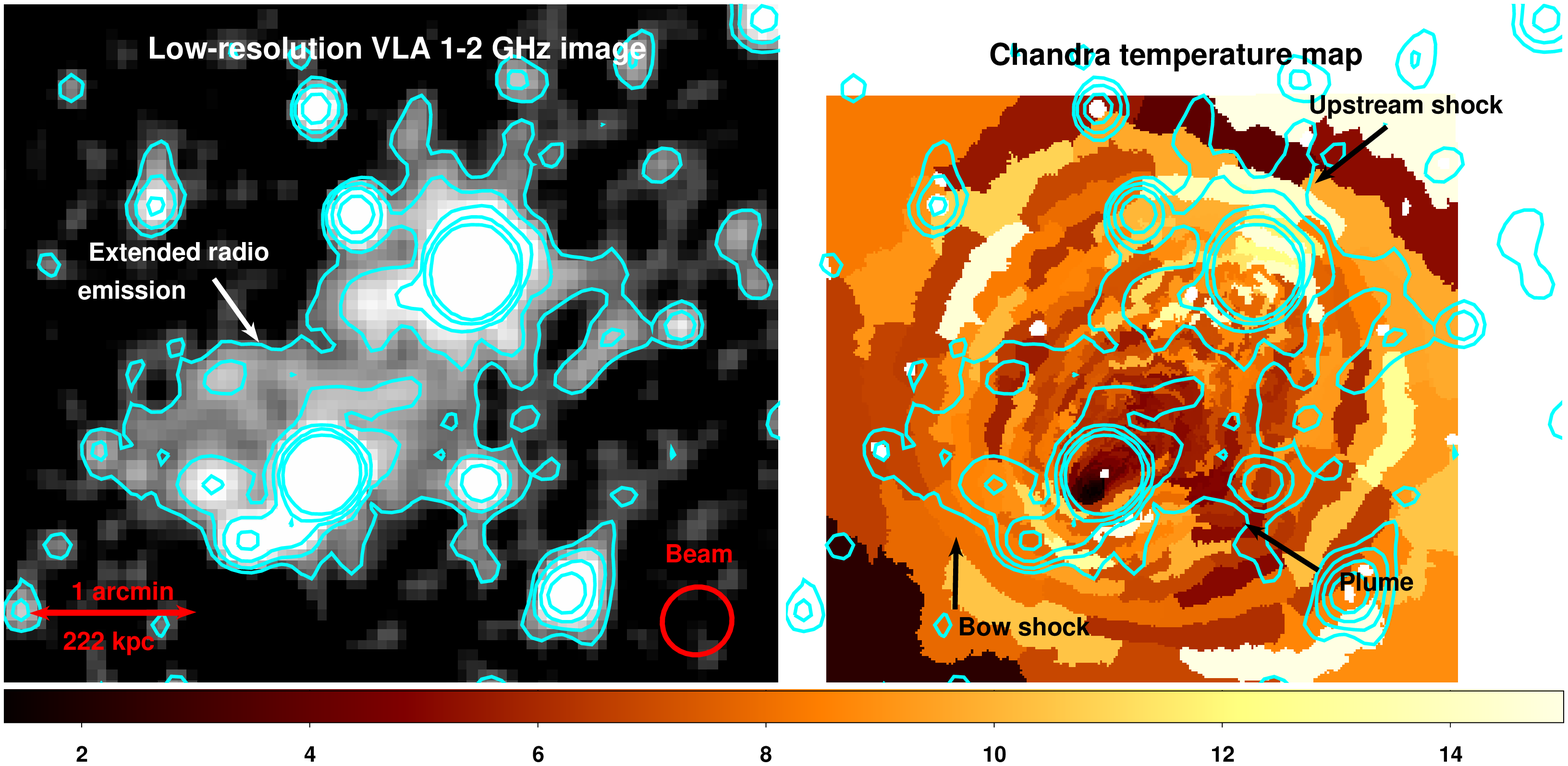}
\end{minipage}
\caption[]{\textbf{Left:} Same low-resolution $1-2$ GHz image as in Fig. \ref{fig3_large}, but zoomed-in onto the cluster on the same scale as Fig. \ref{fig2}. The five contours levels are drawn at $[1,2,4,...]\times3\sigma_{\rm rms}$, where $\sigma_{\rm rms}\sim12\mu$Jy beam$^{-1}$. The beam size is $14.1\times13.5$ arcsec$^2$. \textbf{Right:} \emph{Chandra} projected temperature map from \citet{Rus2012423}. Same contours shown as in the left panel. The colour bar reflects the temperature in keV units.}
\label{fig3}
\end{figure*}

The bottom-left panel of Fig. \ref{fig1} shows the deep 16-hour $1-2$ GHz VLA observations of A2146 at high-resolution. This image was produced by oversampling the beam by a factor of at least 3. It was therefore designed to maximise the resolution and bring out point sources. This image reveals a multitude of radio point sources in the vicinity of the cluster. As in the GMRT 325 MHz image (lower-right panel of Fig. \ref{fig1}), we detect both A2146-A and the northern bright source, but we also detect A2146-B at a faint 3$\sigma_{\rm rms}$ level, where $\sigma_{\rm rms}$ is the local noise level. In addition, both A2146-A and the northern bright source appear to be extended compared to the beam size (shown in the bottom corner of the bottom-left panel of Fig. \ref{fig1}).

In Fig. \ref{fig2}, we show the same VLA image, along with the Subaru R-band image, but we highlight in red and blue the 68 spectroscopically confirmed cluster members found by \citet{Whi2015453}, as well as those from the strong lensing analysis by Coleman et al. (submitted). We identify all cluster members that have a 3$\sigma_{\rm rms}$ radio detection coincident with the galaxy center. Nine of the 68 spectroscopically confirmed cluster members are detected at this level (shown with the blue regions), but we note that in addition to these 9 sources, several other galaxies are detected at a $2\sigma_{\rm rms}$ level. In the discussion, we focuss only of the 3$\sigma_{\rm rms}$ detections.

When producing the VLA images shown in Figs. \ref{fig1} and \ref{fig2}, the parameters in the \textsc{clean} task were chosen such that it would maximise the resolution. However, when imaged individually, both the VLA C-array and D-array datasets reveal the presence of radio emission that extend well beyond the beam in these images, indicating the presence of an additional faint and extended radio component associated with the cluster. 

To demonstrate the existence of this radio component, in Fig. \ref{fig3_large}, we show the large-scale VLA image (B-array, C-array and D-rray), but at low-resolution with a beam size of $14.1\times13.5$ arcsec$^2$ and position angle (PA) of $-27.5^o$. This image was created by
applying an outertaper of 15$''$, corresponding to a
scale of 55 kpc, a \textsc{robust} factor set to $-0.25$ and a cell size of 4$''$. It highlights the presence of a faint radio structure that extends on the scale of the cluster, detected above a $3\sigma_{\rm rms}$ level where $\sigma_{\rm rms}\sim12\mu$Jy beam$^{-1}$.

\begin{figure*}
\centering
\begin{minipage}[c]{0.9\linewidth}
\centering \includegraphics[width=\linewidth]{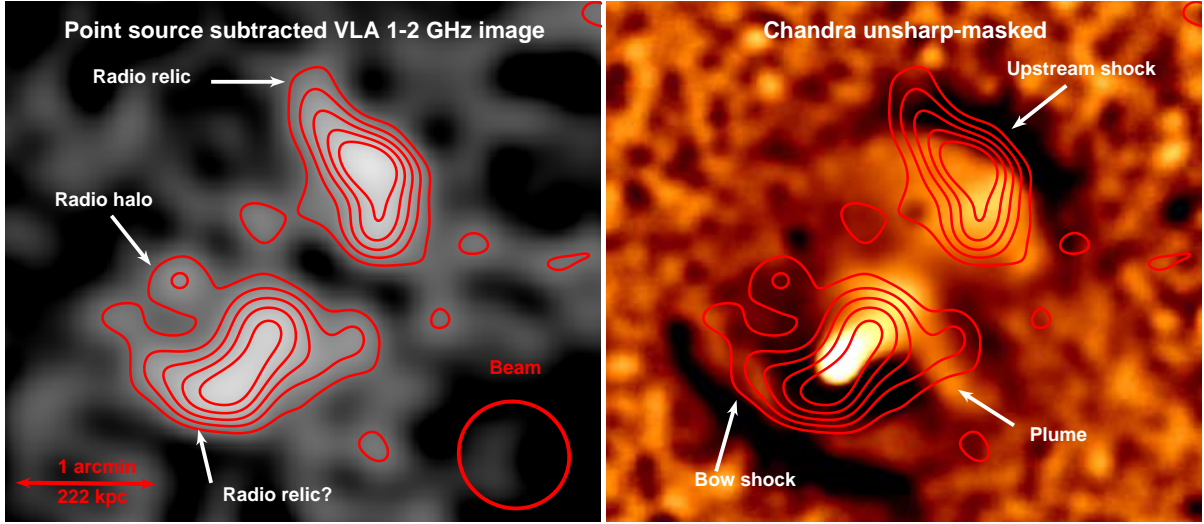}
\end{minipage}
\caption[]{\textbf{Left:} Point source subtracted $1-2$ GHz image on the same scale as Figs. \ref{fig2} and \ref{fig3}. The 6 contours levels are drawn at $[3,5,7,...]\times\sigma_{\rm rms}$, where $\sigma_{\rm rms}\sim23\mu$Jy beam$^{-1}$. The beam size is $27.7\times29.0$ arcsec$^2$. \textbf{Right:} Unsharp-masked \emph{Chandra} image in the $0.3-7.0$ keV band.}
\label{fig3_noptsources}
\end{figure*}

In Fig. \ref{fig3}, we show the same low-resolution VLA image, but zoomed-in onto the cluster on the scale of Fig. \ref{fig2}. We also show the projected temperature map from \citet{Rus2012423}, created by extracting the temperature in binned regions reaching a signal-to-noise ratio of 32 ($\approx1000$ X-ray counts). 

Finally, in Fig. \ref{fig3_noptsources} we show a point source subtracted VLA image. Here, we first create a point source only image by running a \textsc{clean} task and applying a uvrange cutoff ($>$1k$\lambda$) while using a normal briggs \textsc{robust} factor set to $0$. We then subtract this measurement set from the original dataset using the task \textsc{uvsub}. Finally, we image the point source subtracted dataset while using a \textsc{robust} factor set to $0$, no inner uvrange cutoff and a uvtaper of 30$''$. In all cases clean boxes were used. This image reveals that the extended radio emission is actually composed of two components, a first extended component associated with the subcluster core and a second, narrower component associated with the upstream shock. We also show the $0.3-7.0$ keV unsharp-masked \emph{Chandra} image, created by subtracting the X-ray images smoothed with a Gaussian $\sigma$ of 5 arcsec and 20 arcsec, and then dividing the result by the sum of the two. This X-ray image shows nicely different features, in particular the two shock fronts as well as the plume to the south West. It also shows that the diffuse radio structures are strongly bounded by the shock fronts. We discuss these features further in Section 4.

\section{Discussion}

\begin{table*}
\caption[]{Radio emission from A2146 cluster members. The type refers to compact (C) or extended (E). We also indicate in column 6 the flux density when detected at 325 MHz with the GMRT \citep{Rus2011417} and at 150 MHz with the TGSS. }
\hspace{-0.31in}
\resizebox{18cm}{!} {
\begin{tabular}{lccccc}
\hline
Source & RA & DEC & Type & S$_{\rm 1-2 GHz}$ (mJy) & Notes \\
\hline
S1 &  15:56:19.157 &  +66:20:25.860 & C & 0.26$\pm0.05$ &  \\
S2 &  15:56:16.549 &  +66:20:33.260 & C & 0.18$\pm0.03$ &   \\
S3 &  15:56:14.418 &  +66:20:40.350 & C & 0.05$\pm0.01$ &   \\
S4 (A2146-A; BCG) &  15:56:14.002 &  +66:20:53.220 & E & 15.6$\pm3.5$ & $S_{\rm 325 MHz}=47\pm5$ mJy;  $S_{\rm 150 MHz}=29\pm8$ mJy; $S_{\rm 5 GHz}=6.6\pm0.3$ mJy\\
S5 &  15:56:02.166 &  +66:21:47.770 & C & 0.08$\pm0.02$ &    \\
S6 &  15:55:55.224 &  +66:21:48.819 & C & 0.06$\pm0.02$ &    \\
S7 (northern source)&  15:56:03.933 &  +66:22:14.410 & E & 36.6$\pm4.0$ &    $S_{\rm 325 MHz}=93\pm9$ mJy; $S_{\rm 150 MHz}=183\pm21$ mJy; $S_{\rm 5 GHz}=8.9\pm0.5$ mJy\\
S8 (A2146-B; BCG) &  15:56:00.149 &  +66:22:23.940 & C & 0.08$\pm0.02$ &    \\
S9 &  15:56:11.734 &  +66:22:37.690 & C & 0.6$\pm0.1$ &    \\
\hline
\end{tabular}}
\label{tab2}
\end{table*}

\subsection{Compact radio sources}

The original GMRT image at 325 MHz revealed the presence of two unresolved point sources: the BCG belonging to the subcluster core (A2146-A) and the northern bright radio source labelled as $radio~source$ in Fig. \ref{fig1}. At 325 MHz, the first has an integrated
flux of $S_{\rm 325 MHz}=47\pm5$ mJy, while the latter has
an integrated flux of $S_{\rm 325 MHz}=93\pm9$ mJy \citep{Rus2011417}.

The high-resolution $1-2$ GHz image shown in Fig. \ref{fig2} also reveals the presence of A2146-A ($S_{\rm 1-2 GHz}=15.6\pm3.5$ mJy) and the northern bright radio source ($S_{\rm 1-2 GHz}=36.6\pm4.0$ mJy), in agreement with
the values reported by the 1.4 GHz NRAO VLA Sky Survey catalogue
\citep[NVSS;][]{Con1998115}. However, our new image shows that both of these sources are extended on scales of 37 kpc (10 arcsec; A2146-A) and 63 kpc (17 arcsec; northern radio source). 

Both of these sources are not detected in the VLA Low Frequency Sky Survey \citep[VLSS, rms$\sim0.1$ Jy beam$^{-1}$;][]{Coh2007134}. However, the reprocessed 150 MHz images obtained with the alternative TIFR GMRT Sky Survey \citep[TGSS\footnote{http://tgssadr.strw.leidenuniv.nl/doku.php};][]{Int2016598} detects both A2146-A ($S_{\rm 150 MHz}=29\pm8$ mJy) and the northern bright radio source ($S_{\rm 150 MHz}=183\pm21$ mJy), although the first falls below the catalogue detection limit. Interestingly, the TGSS 150 MHz image also reveals that the northern radio source is extended and surrounded by faint $2\sigma$ radio emission. We include the values for the TGSS in Table \ref{tab2}. Finally, we include values at 5 GHz in Table \ref{tab2} for A2146-A and the northern source, based on archival VLA observations (private communication, see also Hogan et al. 2015)\nocite{Hog2015453}. We note that the 408 MHz images taken with the Westerbork Northern Sky Survey \citep[WENSS;][]{Has198247} detects the two radio sources but barely resolves them. 

\begin{figure}
\centering
\begin{minipage}[c]{0.99\linewidth}
\centering \includegraphics[width=\linewidth]{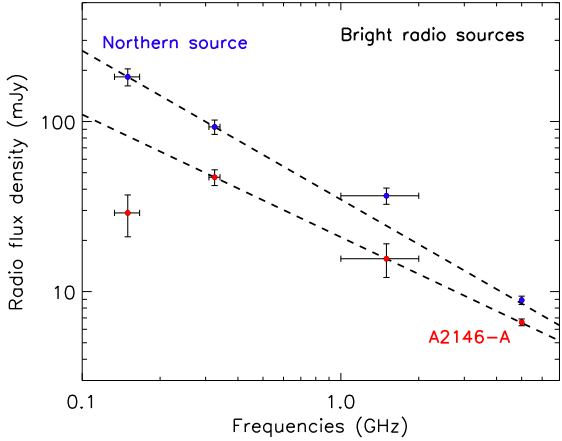}
\end{minipage}
\caption[]{\textbf{Left:} Shown are the flux densities of the two bright radio sources in A2146: A2146-A (red points) and the northern bright source (blue points). The dashed lines highlight the expected fluxes when extrapolating from the measurements at 325 MHz and 1.4 GHz for A2146-A and from 150 MHz and 325 MHz for the northern bright source. }
\label{fig5}
\end{figure}

In Fig. \ref{fig5}, we show the spectral energy distribution of A2146-A (red points) and the northern bright source (blue points). The dashed lines show the expected flux density values for spectral indexes calculated between 325 MHz and 1.4 GHz for A2146-A ($\alpha=0.7\pm0.2$) and between 150 MHz and
325 MHz values for for the northern source ($\alpha=0.9\pm0.1$). This figure highlights two important results, that the spectrum of the northern source rolls over at high frequencies, indicative of synchrotron spectral ageing, and that the spectrum of A2146-A turns over at $\sim200-300$ MHz. Such a turnover is characteristic of a Compton Steep Spectrum (CSS) source, which are usually associated with powerful compact radio sources in their early stages of formation. We note that \citet{Ode1998110} report a correlation between frequency turnover and linear projected size for CSS and gigahertz peaked-spectrum (GPS). For a turnover frequency of $\sim200-300$ MHz, the correlation predicts that A2146-A should extend no more than $\approx30-40$ kpc, which is just within the linear sizes we find for A2146-A. 

The VLA image shown in Fig. \ref{fig2} also reveals the presence of several other radio point sources not previously known. Nine of these radio sources coincide with known cluster members (labelled S1 to S9 in Fig. \ref{fig2}), as identified by \citet{Whi2015453} and Coleman et al. (submitted). The radio fluxes of these galaxies are shown in Table \ref{tab2}. The only radio source that is also clearly detected at X-ray wavelengths is A2146-A (see Fig. \ref{fig1}). 

A detailed analysis of the population of radio sources in A2146 will be conducted in a future paper (Hlavacek-Larrondo et al. in prep). However, we note that A2146 appears to harbour an unusually large population of radio point sources (see King et al. in prep), which may indicate that cluster mergers strongly affect their population of galaxies either through star formation or AGN emission that shines at radio wavelengths \citep[e.g][]{Mil2003125,Pog2004601}.

\subsection{Detection of extended radio emission}

The original GMRT image at
325 MHz did not reveal the presence of a radio halo. This is not surprising considering that the GMRT is known to have problems in
achieving high dynamic range images. Indeed, remaining
artefacts in the form of ripples
surrounding the two point sources could be seen in the original image (see lower-right panel of Fig. \ref{fig1}), which increased the noise level
to 250 $\mu$Jy per beam in the inner regions. For a 1 Mpc source, \citet{Rus2011417} found that the
3$\sigma$ upper limit at 325 MHz for extended radio emission was 13$~$mJy
determined as $3\times{\rm
  rms}\times\sqrt{\rm{source~area/beam~area}}$. Assuming a spectral index
of $\alpha=1$ defined such that $S_\nu\propto\nu^{-\alpha}$,
\citet{Rus2011417} further estimated that the 3$\sigma$ upper limit
was 3$~$mJy ($\sim5.0\times10^{23}$ W Hz$^{-1}$) at $1.4$ GHz.

Our new radio image reaches a noise level of $\sim10\mu$Jy beam$^{-1}$, which is an order of magnitude
deeper than the GMRT image. This has allowed us to discover that A2146 hosts a large extended radio structure that may harbour multiple components, one associated with the subcluster core to the south-east and another associated with the upstream shock to the north-west. As shown in Figs. \ref{fig3} and \ref{fig3_noptsources}, the radio emission is
 centrally located within the cluster, has a morphology that extends along the merger axis of the merger (north-west to south-east) with no clear association to individual galaxies. Considering both components, it has a largest linear size of $\sim850$ kpc and sharp edges that coincide with
both the bow and upstream X-ray shock fronts. Such edges have also been observed in other merging clusters including the Bullet cluster (see Shimwell et al. 2014).

We note that the literature has reported the existence of several elongated, highly-polarized arc-like radio structures known as radio relics \citep[see for a review][]{Bru201423}. These relics are thought to trace the outward going shock fronts produced by the cluster mergers \citep[e.g.][and references therein]{Enb1998332}. One possible solution involves Diffusive Shock Acceleration (DSA) in which electrons from the thermal pool of particles go through several collisions across the shock fronts \citep[e.g.][]{Dru198346}. However, DSA is known to have several issues and \citet{vanW2017Nat} recently argued that relics originate from the re-acceleration of fossil relativistic electrons from past active galactic nuclei outflows, based on the discovery of a direct connection between a radio relic and a radio galaxy in the merging galaxy cluster Abell 3411-3412. Several clusters with X-ray detected shock fronts host such relics coincident with the shocks \citep[e.g.][]{Mac2011728,Ogr2013433}.

In the case of A2146, the elongated morphology of the north-west radio component, as well as its strong association with the upstream shock as shown in Figs. \ref{fig3} and \ref{fig3_noptsources}, leads us to surmise that this structure is most likely a radio relic \citep{Mar2010,Mac2011728,vanW2016818}. Based on our current understanding of radio relics from \citet{vanW2017Nat}, the fossil electrons from this relic may have originated from the bright northern radio source in A2146 (labelled as S7 in Fig. 2), but higher resolution images are needed to confirm this association. 

The origin of the south-east radio component is however more uncertain. Figs. 1 and 4 show that A2146 harbours a cool core that has survived the merger. This cool core also appears to coincide with the south-east radio component, implying that the radio emission seen to the south-east could originate from a classical mini-halo associated with the cool core. However, given the clear nature of the merger in A2146, the extended morphology  of the south-east radio component ($\sim300$ kpc) and the fact that many halos (as opposed to mini-halos) appear to align with their subcluster cores, we conclude that the south-east radio component is most likely not associated with a mini-halo. Instead, we argue that it is most likely associated with a radio halo either bounded by the bow shock \citep[as often seen, e.g][]{Shi2014440} or composed of a halo and a second relic that coincides with the bow shock \citep[e.g][]{vanW2010Sci}. It could also consist of an unusual large box-like relic, such as the one seen in A2256 \citep{Owe2014794}, but this would imply that the second relic extends significantly in the downstream region of the shock front given that A2146 is undergoing a merger in the plane of the sky. We therefore favour the interpretation in which the south-east radio structure is composed of a halo or a halo and relic. Since relics are known to be highly-polarized compared to halos, resolved polarization measurements could provide valuable information, but we note that the intrumental set-up of the VLA observations were not designed to do these measurements. The resolution of the current VLA images are also not sufficient to determine if the radio structure seen to the south-east actually habours multiple components. In summary, future observations will be required to determine the precise nature of the south-eastern component. For now, given its morphology and association with the subcluster core, we consider this structure to be a radio halo.

As shown in Fig. \ref{fig3_noptsources}, the radio halo (south-eastern component) and relic (north-western component) are very faint and were only detectable through the very deep VLA images. To estimate the radio powers, we measure the total emission detected above $3\sigma_{\rm rms}$ in Fig. \ref{fig3_noptsources}. We find that the halo has a flux density of $S_{\rm 1.4 GHz}=1.5$ mJy, whereas the relic has a flux density of $S_{\rm 1.4 GHz}=1.1$ mJy. We then compute the $k-$corrected radio power using $P_{\rm 1.4 GHz}=4{\pi}S_{\rm 1.4 GHz}D_{\rm L}^2(1+z)^{-(-\alpha+1)}$ \citep[see][]{vanW2014786}. For the halo, we use the derived spectral index of $\alpha\approx1$ where $S\propto\nu^{-\alpha}$, see below, but stress that even if we used a value of $\alpha\approx2$, the flux would only change at the 20 per cent level. For the relic, we use the derived spectral index of $\alpha\approx2$. Following \citet{Cas2013777}, we estimate the error on the flux density, and thus the radio power, as
\begin{equation}
\sigma_{\rm S_{\rm{Halo}}} = \sqrt{(\sigma_{\rm cal}S_{\rm halo})^2 + (\rm rms\sqrt{N_{\rm beam}})^2 + \sigma_{\rm sub}^2}
\label{eq1}
\end{equation}
which considers the uncertainties on the flux density scale ($\sigma_{\rm cal}$), typically on the order of 5 per cent, and the noise level of the image (rms$=23\mu$Jy beam$^{-1}$) weighed by the number of beams within the structures. We do not consider $\sigma_{\rm sub}$, the uncertainty when removing the point sources, since the point sources have already been removed in Fig. \ref{fig3_noptsources}. Overall, we find that the point-source subtracted $k-$corrected radio power of the halo (south-eastern component) is $P_{\rm 1.4 GHz}=2.4\pm0.2\times10^{23}$ W Hz$^{-1}$ and for the relic (north-western component) its $P_{\rm 1.4 GHz}=2.2\pm0.2\times10^{23}$ W Hz$^{-1}$.

A precise calculation of the spectral indexes can be determined when radio structures have been detected at multiple frequencies (see Thierbach et al. 2003 for the Coma cluster; Komissarov $\&$ Gubanov for A1914; Bacchi et al. for A754; Feretti et al. for A2319). In the case of A2146, we have only detected the extended radio emission at $1-2$ GHz and can therefore only obtain a rough estimate of the spectral indexes based on this image. In this case,  we use the point source subtracted image of Fig. \ref{fig3_noptsources} and we only consider data points detected above $3\sigma_{\rm rms}$, where $\sigma_{\rm rms}$ is the local noise level. We estimate $\alpha$ by dividing the total integrated flux of the second Taylor term image (.tt1 image produced with the \textsc{clean} task) with the the total integrated flux of the total-intensity image (.tt0 image). This allows us to obtain an estimate of the average spectral index integrated over the entire radio structure. We find a value of $\alpha=1.2\pm0.1$ for the halo (south-east component) and $\alpha=2.3\pm0.3$ for the relic (north-west component), as well as an average value of $\alpha=1.7\pm0.1$ when considering both components. We note that the uncertainty quoted on these values reflects only the uncertainty of the fluxes measured via the .tt1 and .tt0 images. Considering that we are estimating the spectral index based only on the detection of the halo in a very narrow frequency range $1-2$ GHz, the true uncertainty is likely significantly higher. We therefore stress that these are only rough estimates and that a detection at other frequencies is needed to further constrain the spectral index measurements of the radio emission in A2146.

Nonetheless, the derived spectral index of the relic from the VLA $1-2$ GHz image ($\alpha\approx2.3$) is roughly consistent with the expected injection spectral index ($\alpha_{\rm inj}$) predicted by DSA, where 
\begin{equation}
\alpha_{\rm inj}=\frac{\mathcal{M}^2+1}{\mathcal{M}^2-1}-\frac{1}{2}.
\label{eq2}
\end{equation}
Indeed, based on deep \emph{Chandra} observations, \citet[][]{Rus2012423} estimate that the Mach number for the upstream shock is $\mathcal{M}=1.6\pm0.1$, which would imply that $\alpha_{\rm inj}\approx1.8$. However, we stress once more that a better constraint on the spectral index by detecting the relic at other frequencies beyond $1-2$ GHz is needed to test this further.

\begin{figure}
\centering
\begin{minipage}[c]{0.99\linewidth}
\centering \includegraphics[width=\linewidth]{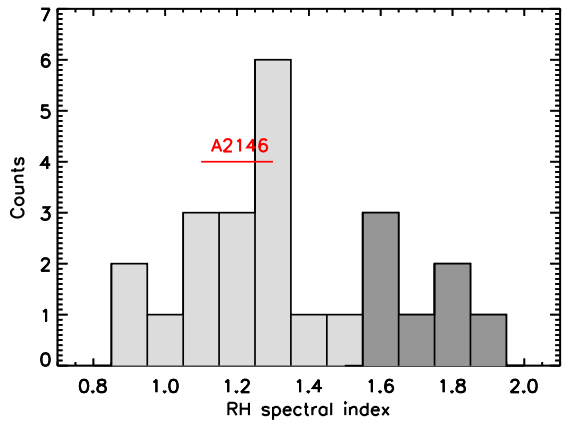}
\end{minipage}
\caption[]{Spectral index measurements ($\alpha$) of radio halos, taken from Feretti et al. (2012); Venturi et al. (2013); Bonafede et al. (2014a, 2014b); Knowles et al. (2016)\nocite{Fer201220,Ven2013551,Bon2014444,Bon2014785,Kno2016459}. The dark grey area shows the USSRH sources with $\alpha>1.6$. Also highlighted is the value of $\alpha$ in A2146, estimated to be roughly $1.2\pm0.1$ based on the VLA $1-2$ GHz image. }
\label{fig4}
\end{figure}

\subsection{Properties of the radio halo}

In Fig. \ref{fig4}, we show a compilation of radio halo spectral index measurements from the literature (Feretti et al. 2012; Venturi et al. 2013; Bonafede et al. 2014a, 2014b; Knowles et al. 2016)\nocite{Fer201220,Ven2013551,Bon2014444,Bon2014785,Kno2016459}. The values are total spectral index values for straight spectra, or low frequency spectral index values for curved spectra. As shown, most halos have spectral index measurements of $\alpha\approx1-1.5$, but some, known as ultrasteep spectrum radio halos (USSRHs) and shown in dark grey in Fig. \ref{fig4} have $\alpha\approx1.6-1.9$. These are thought to be the older counterparts of halos, in which the population of electrons has aged, causing the spectra to steepen even further \citep[e.g.][]{Ven2013551}. The value derived for the spectral index of the halo (south-eastern component) in A2146 (roughly $\alpha_{\rm 1-2GHz}\approx1.2$; shown in red in Fig. \ref{fig4}) does not appear to be consistent with the aged population of halos. We also note that the non-detection of the halo at 325 MHz with the GMRT also implies that the halo must have a low value of the spectral index, roughly $\alpha\approx1.0$.

\citet{Don2013429} argue that the properties of halos in clusters of galaxies vary depending on the stage of the merger. Based on high-resolution magnetohydrodynamic simulations of a $1.5\times10^{15}M_{\rm \odot}$ cluster with a mass ratio of $1:8$, the authors found that both X-ray and radio luminosities increase during the infall phase ($<0.1$ Gyr after core passage), and that radio emission is localized within the shock fronts. Roughly $\approx0.1-0.5$ Gyr after core passage, the X-ray emission declines due to the decrease in density, while the radio emission continues to increase as turbulence drives the re-acceleration of particules throughout the cluster volume. Finally, $>0.5$ Gyr after core passage, the radio emission fades significantly and becomes offset from the primary core of the merger, essentially entering a \emph{radio-off} state. In addition to variations of the radio intensity, \citet{Don2013429} predict that the spectral index of the halo will vary with time, from flat spectra early on ($\alpha\approx1.0$), down to very steep spectra at late times ($\alpha>1.5$). 

In the case of A2146, Fig. \ref{fig3_noptsources} shows that the halo is strongly confined within the bow shock front such that the merger-induced turbulence has not yet reached the entire cluster. We note in particular that the halo appears to be more extended along the eastern direction, even though the bow shock extends to the west, see Fig. \ref{fig3_noptsources}. This is intriguing given that the temperature map in Fig. \ref{fig3} shows that there is more mixing (i.e. turbulence) on the eastern side of the bow shock, as opposed to the western side. The halo may also extend to the southern X-ray plume. 

In addition, the radio power of the halo is among the lowest detected thus far \citep[$P_{\rm 1.4 GHz}=2.4\pm0.2\times10^{23}$ W Hz$^{-1}$; e.g.][]{Kno2016459}, and our rough estimate of the spectral index based on the VLA image is consistent with the younger population of halos ($\alpha<1.5$). All of these observations are consistent with the halo being in a very early stage of formation, roughly $\approx0.3$ Gyr after core passage according to the simulations of \citet[][]{Don2013429}, although A2146 is thought to be a merger with a much smaller mass ratio of $4-3:1$ \citep{Rus2010406,Whi2015453} compared to the cluster in the simulations. Interestingly, this falls within the age estimate obtained by \citet{Whi2015453}, who used spectroscopic measurements of cluster members to establish the dynamical state of A2146. Based on a two-body dynamical model, the authors estimated that the time-scale after first core passage must be on the order of $\approx 0.24-0.28$ Gyr. We therefore conclude that the radio halo in A2146 is not only among the faintest ever detected, but that its properties are also consistent with the halo being in its early stages of formation. This makes the halo in A2146 extremely interesting to study, since it allows us to probe the complete evolutionary stages of halos.

\subsection{Implications for radio halos}

Radio halos are not ubiquitous in galaxy clusters. Currently, the literature reports the existence of $\approx40$ halos \citep{Gia2017841}, all of which are identified in massive clusters ($M>>10^{14}$M$_{\rm \odot}$). Halos are thought to originate from the re-acceleration of in situ relativistic particles by turbulence generated from the cluster merger \citep[e.g.][]{Lia2000544,Cas2006369}. Radio halos may also be produced by inelastic hadronic collisions between cosmic-ray protons and thermal protons \citep[see e.g.][]{Den1980239}, but this scenario is less favoured because of the lack of predicted gamma-ray emission and the existence of USSRHs. For the latter, the hadronic models predict that the relativistic energy would exceed the thermal energy, which is implausible. 

\begin{figure*}
\centering
\begin{minipage}[c]{0.32\linewidth}
\centering \includegraphics[width=\linewidth]{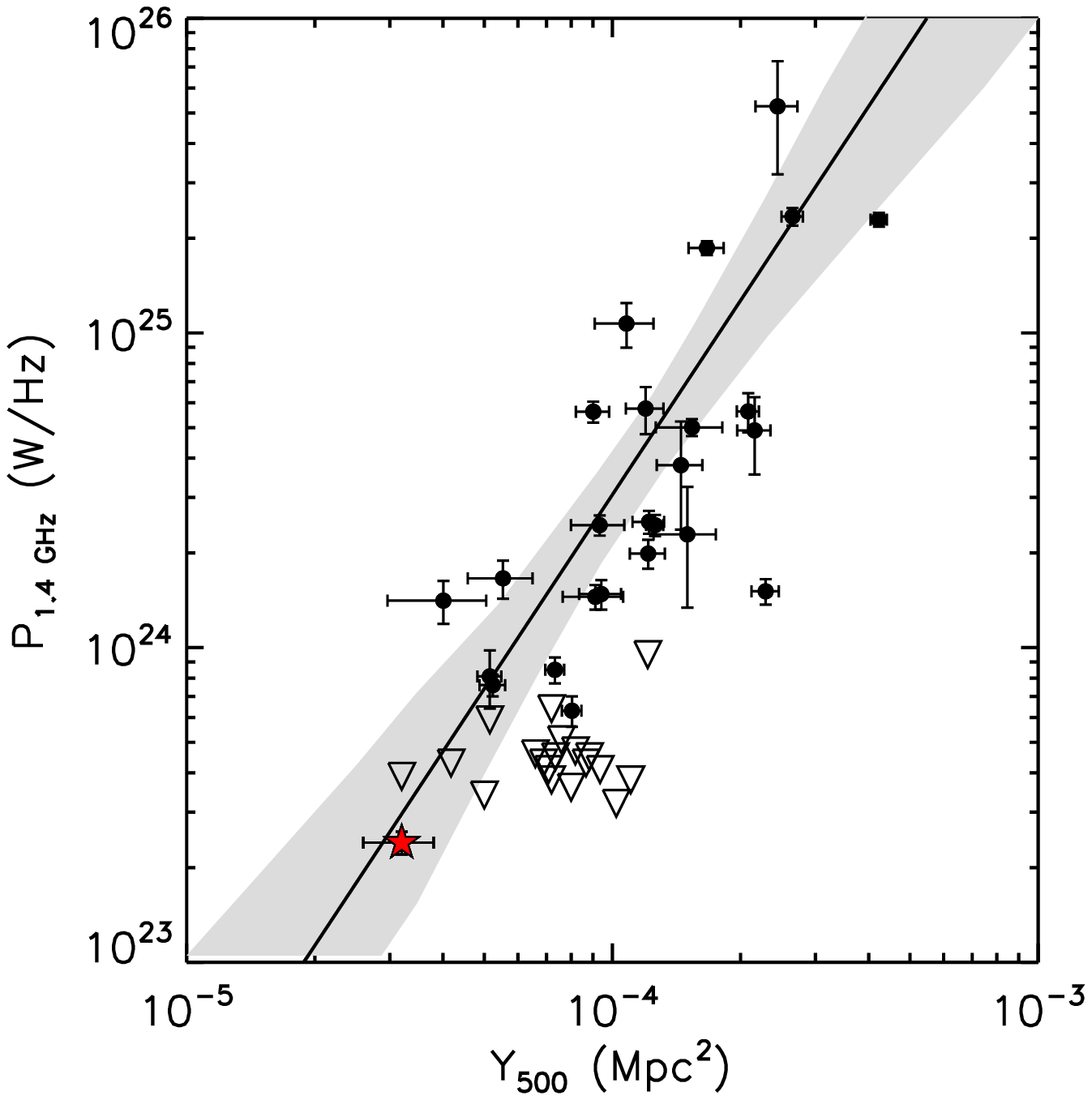}
\end{minipage}
\begin{minipage}[c]{0.32\linewidth}
\centering \includegraphics[width=\linewidth]{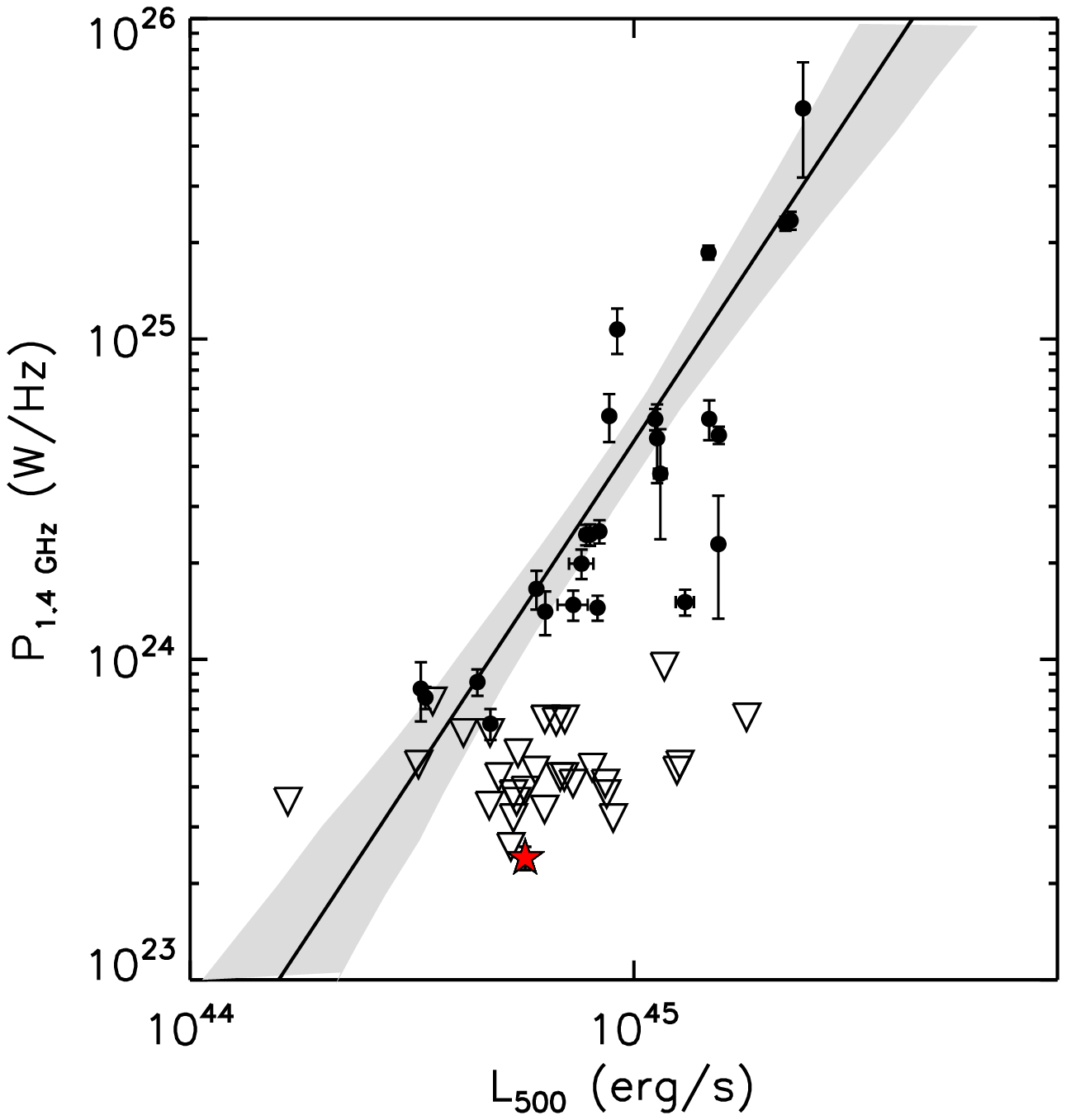}
\end{minipage}
\begin{minipage}[c]{0.32\linewidth}
\centering \includegraphics[width=\linewidth]{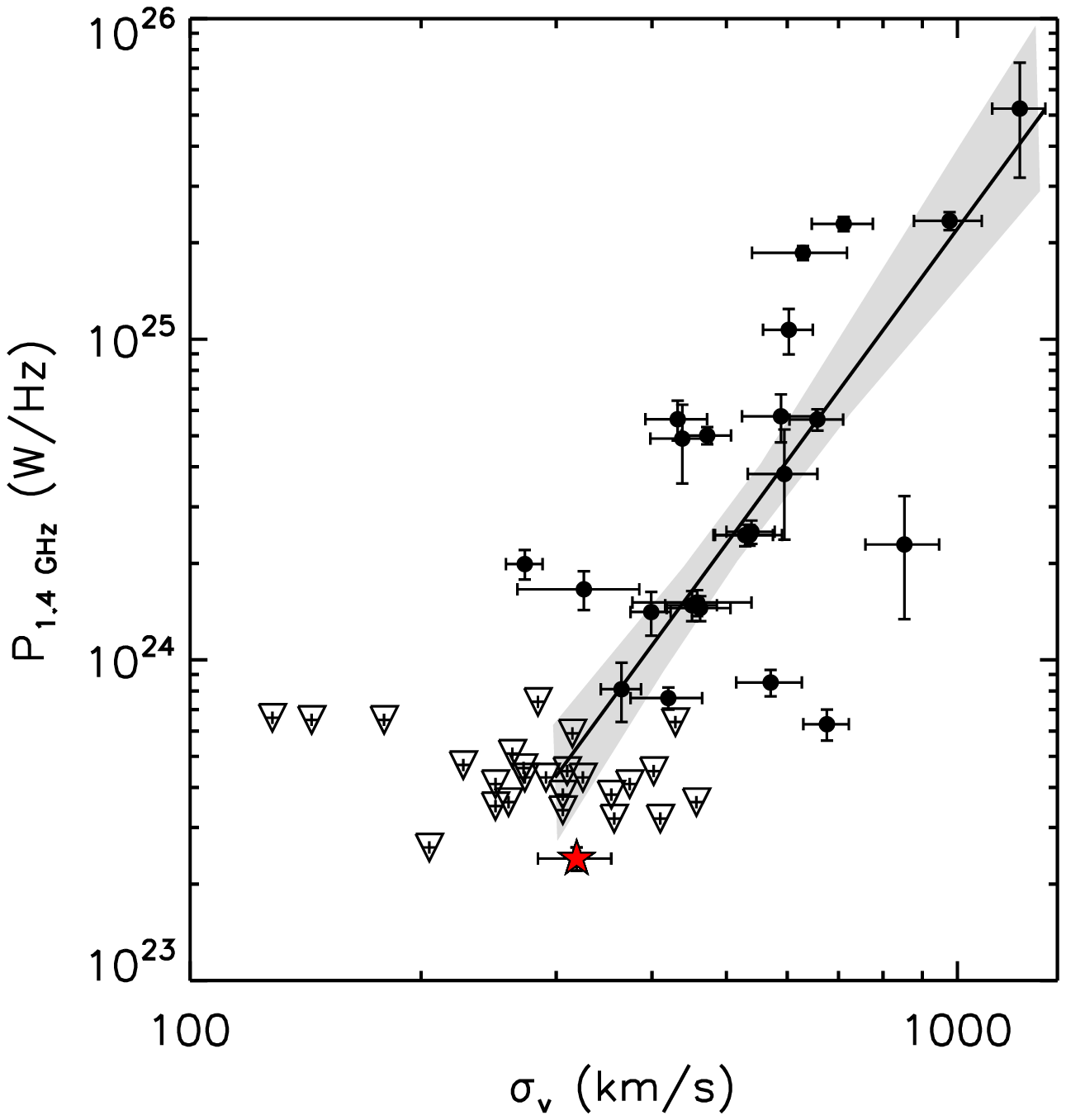}
\end{minipage}
\caption[]{\textbf{Left: }Correlation of the radio halo power at 1.4 GHz ($P_{\rm 1.4 GHz}$) as a function of the integrated SZ signal ($Y_{500}$) of the cluster within $R_{500}$ \citep{Cas2013777}. USSRHs lie to the right of the correlation. The shaded region shows the 95 per cent confidence level. \textbf{Middle: }Correlation of the radio halo power at 1.4 GHz ($P_{\rm 1.4 GHz}$) as a function of the X-ray luminosity ($L_{500}$) of the cluster within $R_{500}$, also from \citep{Cas2013777}. The shaded region also shows the 95 per cent confidence level. \textbf{Right: }Correlation of the radio halo power at 1.4 GHz ($P_{\rm 1.4 GHz}$) as a function of the turbulent velocity measured from the amplitude of density fluctuations ($\sigma_{\rm v}$) for 51 clusters \citep{Eck2017843}. The shaded region shows the 68 per cent confidence region. The halo in A2146 is highlighted with the red star in all three panels. }
\label{fig6}
\end{figure*}

Radio haloes are known to exhibit empirical correlations between radio halo power and various thermal cluster properties, including X-ray luminosity, gas temperature, cluster mass and the cluster integrated Sunyaev-Zel'dovich (SZ) signal \citep[see][for a recent re-visitation of the relations]{Cas2013777}. Several studies confirm that the correlations exhibit a bimodal distribution, with the X-ray luminous or massive clusters either falling on the correlation or a factor $2-3$ below, known as the \emph{radio-off} state (Venturi et al. 2007, 2008). In the case of A2146, because of the non-detection of the halo in A2146 with the GMRT, \citet{Rus2011417} argued that A2146 may not be massive enough to generate a luminous halo, explaining why the cluster falls well below the correlation even if strongly merging. Low-mass systems are expected to generate less turbulence, producing less luminous halos. If this is the case, radio halos might therefore be significantly more numerous than thought, but our current view is limited by the sensitivity of radio telescopes (e.g. GMRT, VLA).

In the left panel of Fig. \ref{fig6}, we illustrate the location of the halo in A2146 compared to other clusters for the 1.4 GHz radio halo power versus integrated SZ signal (Y$_{500}$) within a radius at which the average density is 500 times the critical density at the cluster redshift (R$_{500}$) from \citet{Cas2013777}. Although the X-ray luminosity of the cluster might be boosted during the merger phase, we also show in Fig. \ref{fig6} (middle-panel) the $P_{\rm 1.4 GHz}-L_{500}$ diagram. Here, we use the X-ray luminosity within R$_{500}$ derived by \citet[][]{Cas2013777}. We do not show the $P_{\rm 1.4 GHz}-M_{500}$ correlation, since the mass estimate for A2146, given its complex dynamical state, varies significantly from one study to the next: from $M_{500}\approx~3.8\times10^{14}M_{\rm \odot}$ for an SZ estimate with the Planck satellite \cite[][]{Pla2014571}, to $M_{\rm 200}=8.5^{+4.3}_{-4.7}\times10^{14}M_{\rm \odot}$ based on a dynamical analysis of the cluster members \citep[][]{Whi2015453}, to $M_{\rm 200}=1.4^{+0.3}_{-0.4}\times10^{15}M_{\rm \odot}$ for a \emph{Hubble Space Telescope} weak lensing analysis \citep[][]{Kin2016459}. Instead, we focus on the X-ray luminosity and integrated SZ measurement obtained with the Planck satellite (Y$_{500}=0.000032\pm0.000006$ Mpc$^2$), which is consistent with the value obtained with the Arcminute Microkelvin Imager (AMI; Y$_{500}=0.000029\pm0.000014$ Mpc$^2$, private communication; see also Zwart et al. 2008, Rodriguez-Gonzalvez et al. 2011)\nocite{Zwa2008391,Rod2011414}. The left panel of Fig. \ref{fig6} essentially shows that the halo of A2146 is indeed one of the faintest detected thus far. It also appears to be consistent with the $P_{\rm 1.4 GHz}-Y_{500}$ correlation, although offset from the $P_{\rm 1.4 GHz}-L_{500}$ correlation (however, we note that the X-ray luminosity of the cluster might be boosted during the merger phase). We also note that, as argued in Section 4.2, if the halo (south-east component seen in Fig. \ref{fig3_noptsources}) were actually composed of a halo and a relic, the derived radio power of the halo would be even smaller and A2146 would sit even further below the correlations, still making it one of the faintest ever detected.

More recently, \citet{Eck2017843} derived the velocity dispersion of 51 clusters based on the amplitude of density fluctuations, and found that the turbulent velocity of the cluster with radio halos was on average a factor of two higher than those without halos. This result supports the scenario in which halos are driven by turbulent motions. They also find a strong correlation between radio halo power at 1.4 GHz ($P_{\rm 1.4 GHz}$) and the turbulent velocity ($\sigma_{\rm v}$). These authors include A2146 in their study, for which they find that $\sigma_{\rm v}=319\pm5$ km s$^{-1}$. The detection of the candidate halo in A2146 with $P_{\rm 1.4 GHz}=2.4\pm0.2\times10^{23}$ W Hz$^{-1}$ places A2146 within the scatter of the correlation observed by \citet{Eck2017843}, see right panel of Fig. \ref{fig6}.

\section{Concluding remarks}

We have detected a diffuse, radio structure associated with the merging cluster A2146 using deep, multi-configuration VLA observations at $1-2$ GHz (16 hours). The radio structure is faint ($P_{\rm 1.4 GHz, total}=4.6\pm0.3\times10^{23}$ W Hz$^{-1}$), elongated along the merger axis and strongly confined between both the bow and upstream X-ray shock fronts. This structure extends over 850 kpc in length and appears to harbour multiple components, one associated with the upstream shock which we classify as a radio relic, and another associated with the subcluster core which is consistent as being a radio halo bounded by the bow shock. The latter is among the faintest radio halos detected thus far in the bimodal distributions of radio halo power ($P_{\rm 1.4 GHz}$) versus thermal cluster properties (Y$_{500}$, L$_{500}$, M$_{500}$). We note that the halo could actually be composed of a halo and a second relic that coincides with the bow shock, but the current observations do not allow us to distinguish between possibilities. Future observations, in particular via polarization, could provide a definitive answer. However, if this were the case, than the radio halo would be even fainter. 

The spectral index of the halo, derived from the $1-2$ GHz VLA image, as well as its non-detection at 325 MHz with the GMRT imply that the structure must have a low value of the spectral index, with $\alpha\approx1$. The confinement of the halo within the X-ray shock fronts indicates that the merger-induced turbulence has not yet reached the cluster scale and that the halo is most likely in its early stages of formation. The low spectral index and the location of A2146 in the $Y_{500}-P_{\rm 1.4 GHz}$ correlation also suggest that this halo is in its early stages of formation, roughly $\approx0.3$ Gyr after first core passage, consistent with the dynamical analysis of the cluster members. Such low radio power halos provide a unique opportunity to study the physical processes that govern the initial stages of halo formation, making the study of faint radio halos just as important as the study of the luminous radio halos. 

Given that the detection of the extended radio emission in A2146 was only possible through deep VLA observations reaching a noise level of 10$\mu$Jy beam$^{-1}$, i.e. more than an order of magnitude deeper than previous observations, several other radio halos are bound to be discovered with the next generation of radio telescopes such as the Long Wavelength Array (LWA; Taylor et al. 2012)\nocite{Tay2012}, the MWA (Lonsdale et al. 2009; Bowman et al. 2013; Tingay et al. 2013), LOFAR (LOw Frequency ARray, van Haarlem et al. 2013) and the upcoming SKA (Square Kilometer Array, Dewdney et al. 2013). Indeed, these telescopes will not only reach even lower noise levels (e.g. SKA is predicted to reach flux densities as least 10 times fainter), but they will also provide the good uv-coverage at short baselines needed to detect extended diffuse emission in clusters. This will be extremely important in light of the faint halo and relic discovered in A2146, and for any similar structures caught in their early stages of formation.

In summary, our results highlight the need for additional deep observations of clusters to better understand the faint end of the bimodal $P_{\rm 1.4 GHz}$ versus thermal property correlations, and that many of the clusters in the upper-limit region may indeed host faint radio halos.

\section*{Acknowledgments}

JHL is supported by NSERC through the discovery grant and Canada Research Chair programs, as well as FRQNT through the new university researchers start up program. MLGM is supported by NSERC through the NSERC Post-graduate Scholarship Doctoral Program (PGSD). DFB is supported by NSERC, through the summer internship program. We thank Becky Canning for helpful discussions about the point sources in the cluster, as well as the cluster members.

\bibliographystyle{mn2e}
\bibliography{bibli}

\bsp	
\label{lastpage}

\end{document}

%% file: defn.tex
\newcommand{\apgt}{{\raise-.5ex\hbox{$\buildrel>\over\sim$}}}
\newcommand{\aplt}{{\raise-.5ex\hbox{$\buildrel<\over\sim$}}} 
\newcommand{\Mpc}{\rm\; Mpc}

\newcommand{\km}{\rm\; km}

\newcommand{\s}{\rm\; s}

\newcommand{\kmps}{\hbox{$\km\s^{-1}\,$}}

\newcommand{\kmpspMpc}{\hbox{$\kmps\Mpc^{-1}\,$}}

\newcommand{\Omm}{\hbox{$\rm\thinspace \Omega_{m}$}}
\newcommand{\OmL}{\hbox{$\rm\thinspace \Omega_{\Lambda}$}}









%% file: A2146_HlavacekLarrondo.bbl
\begin{thebibliography}{}

\bibitem[\protect\citeauthoryear{{Ackermann}, {Ajello}, {Albert}, {Allafort} \&
  {et al.}}{{Ackermann} et~al.}{2014}]{Ack2014787}
{Ackermann} M.,  {Ajello} M.,  {Albert} A.,  {Allafort} A.,    {et al.} 2014,
  \apj, 787, 18

\bibitem[\protect\citeauthoryear{{Bonafede}, {Govoni}, {Feretti}, {Murgia},
  {Giovannini} \& {Br{\"u}ggen}}{{Bonafede} et~al.}{2011}]{Bon2011530}
{Bonafede} A.,  {Govoni} F.,  {Feretti} L.,  {Murgia} M.,  {Giovannini} G.,
  {Br{\"u}ggen} M.,  2011, \aap, 530, A24

\bibitem[\protect\citeauthoryear{{Bonafede}, {Intema}, {Br{\"u}ggen},
  {Girardi}, {Nonino}, {Kantharia}, {van Weeren} \&
  {R{\"o}ttgering}}{{Bonafede} et~al.}{2014}]{Bon2014785}
{Bonafede} A.,  {Intema} H.~T.,  {Br{\"u}ggen} M.,  {Girardi} M.,  {Nonino} M.,
   {Kantharia} N.,  {van Weeren} R.~J.,    {R{\"o}ttgering} H.~J.~A.,  2014,
  \apj, 785, 1

\bibitem[\protect\citeauthoryear{{Bonafede}, {Intema}, {Br{\"u}ggen},
  {Russell}, {Ogrean}, {Basu}, {Sommer}, {van Weeren}, {Cassano}, {Fabian} \&
  {R{\"o}ttgering}}{{Bonafede} et~al.}{2014}]{Bon2014444}
{Bonafede} A.,  {Intema} H.~T.,  {Br{\"u}ggen} M.,  {Russell} H.~R.,  {Ogrean}
  G.,  {Basu} K.,  {Sommer} M.,  {van Weeren} R.~J.,  {Cassano} R.,  {Fabian}
  A.~C.,    {R{\"o}ttgering} H.~J.~A.,  2014, \mnras, 444, L44

\bibitem[\protect\citeauthoryear{{Brada{\v c}}, {Clowe}, {Gonzalez},
  {Marshall}, {Forman}, {Jones}, {Markevitch}, {Randall}, {Schrabback} \&
  {Zaritsky}}{{Brada{\v c}} et~al.}{2006}]{Bra2006652}
{Brada{\v c}} M.,  {Clowe} D.,  {Gonzalez} A.~H.,  {Marshall} P.,  {Forman} W.,
   {Jones} C.,  {Markevitch} M.,  {Randall} S.,  {Schrabback} T.,    {Zaritsky}
  D.,  2006, \apj, 652, 937

\bibitem[\protect\citeauthoryear{{Briggs}, {Voges}, {Bohringer}, {Edge},
  {Huchra} \& {Briel}}{{Briggs} et~al.}{1996}]{Bri1995}
{Briggs} H.,  {Voges} W.,  {Bohringer} H.,  {Edge} A.~C.,  {Huchra} J.~P.,
  {Briel} U.~G.,  1996, \mnras, 281, 799

\bibitem[\protect\citeauthoryear{{Brown}, {Emerick}, {Rudnick} \&
  {Brunetti}}{{Brown} et~al.}{2011}]{Bro2011740}
{Brown} S.,  {Emerick} A.,  {Rudnick} L.,    {Brunetti} G.,  2011, \apjl, 740,
  L28

\bibitem[\protect\citeauthoryear{{Brunetti} \& {Jones}}{{Brunetti} \&
  {Jones}}{2014}]{Bru201423}
{Brunetti} G.,  {Jones} T.~W.,  2014, International Journal of Modern Physics
  D, 23, 1430007

\bibitem[\protect\citeauthoryear{{Brunetti}, {Setti}, {Feretti} \&
  {Giovannini}}{{Brunetti} et~al.}{2001}]{Bru2001320}
{Brunetti} G.,  {Setti} G.,  {Feretti} L.,    {Giovannini} G.,  2001, \mnras,
  320, 365

\bibitem[\protect\citeauthoryear{{Canning}, {Russell}, {Hatch}, {Fabian},
  {Zabludoff}, {Crawford}, {King}, {McNamara}, {Okamoto} \&
  {Raimundo}}{{Canning} et~al.}{2012}]{Can2012420}
{Canning} R.~E.~A.,  {Russell} H.~R.,  {Hatch} N.~A.,  {Fabian} A.~C.,
  {Zabludoff} A.~I.,  {Crawford} C.~S.,  {King} L.~J.,  {McNamara} B.~R.,
  {Okamoto} S.,    {Raimundo} S.~I.,  2012, \mnras, 420, 2956

\bibitem[\protect\citeauthoryear{{Cassano}, {Brunetti} \& {Setti}}{{Cassano}
  et~al.}{2006}]{Cas2006369}
{Cassano} R.,  {Brunetti} G.,    {Setti} G.,  2006, \mnras, 369, 1577

\bibitem[\protect\citeauthoryear{{Cassano}, {Ettori}, {Brunetti},
  {Giacintucci}, {Pratt}, {Venturi}, {Kale}, {Dolag} \& {Markevitch}}{{Cassano}
  et~al.}{2013}]{Cas2013777}
{Cassano} R.,  {Ettori} S.,  {Brunetti} G.,  {Giacintucci} S.,  {Pratt} G.~W.,
  {Venturi} T.,  {Kale} R.,  {Dolag} K.,    {Markevitch} M.,  2013, \apj, 777,
  141

\bibitem[\protect\citeauthoryear{{Clowe}, {Brada{\v c}}, {Gonzalez},
  {Markevitch}, {Randall}, {Jones} \& {Zaritsky}}{{Clowe}
  et~al.}{2006}]{Clo2006648}
{Clowe} D.,  {Brada{\v c}} M.,  {Gonzalez} A.~H.,  {Markevitch} M.,  {Randall}
  S.~W.,  {Jones} C.,    {Zaritsky} D.,  2006, \apjl, 648, L109

\bibitem[\protect\citeauthoryear{{Clowe}, {Gonzalez} \& {Markevitch}}{{Clowe}
  et~al.}{2004}]{Clo2004604}
{Clowe} D.,  {Gonzalez} A.,    {Markevitch} M.,  2004, \apj, 604, 596

\bibitem[\protect\citeauthoryear{{Cohen}, {Lane}, {Cotton}, {Kassim}, {Lazio},
  {Perley}, {Condon} \& {Erickson}}{{Cohen} et~al.}{2007}]{Coh2007134}
{Cohen} A.~S.,  {Lane} W.~M.,  {Cotton} W.~D.,  {Kassim} N.~E.,  {Lazio}
  T.~J.~W.,  {Perley} R.~A.,  {Condon} J.~J.,    {Erickson} W.~C.,  2007, \aj,
  134, 1245

\bibitem[\protect\citeauthoryear{{Condon}, {Cotton}, {Greisen}, {Yin},
  {Perley}, {Taylor} \& {Broderick}}{{Condon} et~al.}{1998}]{Con1998115}
{Condon} J.~J.,  {Cotton} W.~D.,  {Greisen} E.~W.,  {Yin} Q.~F.,  {Perley}
  R.~A.,  {Taylor} G.~B.,    {Broderick} J.~J.,  1998, \aj, 115, 1693

\bibitem[\protect\citeauthoryear{{Cornwell}, {Golap} \& {Bhatnagar}}{{Cornwell}
  et~al.}{2005}]{Cor2005347}
{Cornwell} T.~J.,  {Golap} K.,    {Bhatnagar} S.,  2005, in {Shopbell} P.,
  {Britton} M.,   {Ebert} R.,  eds, Astronomical Data Analysis Software and
  Systems XIV Vol.~347 of Astronomical Society of the Pacific Conference
  Series, {W Projection: A New Algorithm for Wide Field Imaging with Radio
  Synthesis Arrays}.
p.~86

\bibitem[\protect\citeauthoryear{{Dennison}}{{Dennison}}{1980}]{Den1980239}
{Dennison} B.,  1980, \apjl, 239, L93

\bibitem[\protect\citeauthoryear{{Donnert}, {Dolag}, {Brunetti} \&
  {Cassano}}{{Donnert} et~al.}{2013}]{Don2013429}
{Donnert} J.,  {Dolag} K.,  {Brunetti} G.,    {Cassano} R.,  2013, \mnras, 429,
  3564

\bibitem[\protect\citeauthoryear{{Drury}}{{Drury}}{1983}]{Dru198346}
{Drury} L.~O.,  1983, Reports on Progress in Physics, 46, 973

\bibitem[\protect\citeauthoryear{{Eckert}, {Gaspari}, {Vazza}, {Gastaldello},
  {Tramacere}, {Zimmer}, {Ettori} \& {Paltani}}{{Eckert}
  et~al.}{2017}]{Eck2017843}
{Eckert} D.,  {Gaspari} M.,  {Vazza} F.,  {Gastaldello} F.,  {Tramacere} A.,
  {Zimmer} S.,  {Ettori} S.,    {Paltani} S.,  2017, \apjl, 843, L29

\bibitem[\protect\citeauthoryear{{Ensslin}, {Biermann}, {Klein} \&
  {Kohle}}{{Ensslin} et~al.}{1998}]{Enb1998332}
{Ensslin} T.~A.,  {Biermann} P.~L.,  {Klein} U.,    {Kohle} S.,  1998, \aap,
  332, 395

\bibitem[\protect\citeauthoryear{{Feretti}, {Giovannini}, {Govoni} \&
  {Murgia}}{{Feretti} et~al.}{2012}]{Fer201220}
{Feretti} L.,  {Giovannini} G.,  {Govoni} F.,    {Murgia} M.,  2012, \aapr, 20,
  54

\bibitem[\protect\citeauthoryear{{Giacintucci}, {Markevitch}, {Cassano},
  {Venturi}, {Clarke} \& {Brunetti}}{{Giacintucci} et~al.}{2017}]{Gia2017841}
{Giacintucci} S.,  {Markevitch} M.,  {Cassano} R.,  {Venturi} T.,  {Clarke}
  T.~E.,    {Brunetti} G.,  2017, \apj, 841, 71

\bibitem[\protect\citeauthoryear{{Haslam}, {Salter}, {Stoffel} \&
  {Wilson}}{{Haslam} et~al.}{1982}]{Has198247}
{Haslam} C.~G.~T.,  {Salter} C.~J.,  {Stoffel} H.,    {Wilson} W.~E.,  1982,
  \aaps, 47, 1

\bibitem[\protect\citeauthoryear{{Hogan}, {Edge}, {Hlavacek-Larrondo},
  {Grainge}, {Hamer}, {Mahony}, {Russell}, {Fabian}, {McNamara} \&
  {Wilman}}{{Hogan} et~al.}{2015}]{Hog2015453}
{Hogan} M.~T.,  {Edge} A.~C.,  {Hlavacek-Larrondo} J.,  {Grainge} K.~J.~B.,
  {Hamer} S.~L.,  {Mahony} E.~K.,  {Russell} H.~R.,  {Fabian} A.~C.,
  {McNamara} B.~R.,    {Wilman} R.~J.,  2015, \mnras, 453, 1201

\bibitem[\protect\citeauthoryear{{Intema}, {Jagannathan}, {Mooley} \&
  {Frail}}{{Intema} et~al.}{2017}]{Int2016598}
{Intema} H.~T.,  {Jagannathan} P.,  {Mooley} K.~P.,    {Frail} D.~A.,  2017,
  \aap, 598, A78

\bibitem[\protect\citeauthoryear{{Jeltema} \& {Profumo}}{{Jeltema} \&
  {Profumo}}{2011}]{Jel2011728}
{Jeltema} T.~E.,  {Profumo} S.,  2011, \apj, 728, 53

\bibitem[\protect\citeauthoryear{{King}, {Clowe}, {Coleman}, {Russell},
  {Santana}, {White}, {Canning}, {Deering}, {Fabian}, {Lee}, {Li} \&
  {McNamara}}{{King} et~al.}{2016}]{Kin2016459}
{King} L.~J.,  {Clowe} D.~I.,  {Coleman} J.~E.,  {Russell} H.~R.,  {Santana}
  R.,  {White} J.~A.,  {Canning} R.~E.~A.,  {Deering} N.~J.,  {Fabian} A.~C.,
  {Lee} B.~E.,  {Li} B.,    {McNamara} B.~R.,  2016, \mnras, 459, 517

\bibitem[\protect\citeauthoryear{{Knowles}, {Intema}, {Baker}, {Bharadwaj},
  {Bond}, {Cress}, {Gupta} \& {et al.}}{{Knowles} et~al.}{2016}]{Kno2016459}
{Knowles} K.,  {Intema} H.~T.,  {Baker} A.~J.,  {Bharadwaj} V.,  {Bond} J.~R.,
  {Cress} C.,  {Gupta} N.,    {et al.} 2016, \mnras, 459, 4240

\bibitem[\protect\citeauthoryear{{Liang}, {Hunstead}, {Birkinshaw} \&
  {Andreani}}{{Liang} et~al.}{2000}]{Lia2000544}
{Liang} H.,  {Hunstead} R.~W.,  {Birkinshaw} M.,    {Andreani} P.,  2000, \apj,
  544, 686

\bibitem[\protect\citeauthoryear{{Macario}, {Markevitch}, {Giacintucci},
  {Brunetti}, {Venturi} \& {Murray}}{{Macario} et~al.}{2011}]{Mac2011728}
{Macario} G.,  {Markevitch} M.,  {Giacintucci} S.,  {Brunetti} G.,  {Venturi}
  T.,    {Murray} S.~S.,  2011, \apj, 728, 82

\bibitem[\protect\citeauthoryear{{Markevitch}}{{Markevitch}}{2010}]{Mar2010}
{Markevitch} M.,  2010, ArXiv e-prints, arXiv:1010.3660

\bibitem[\protect\citeauthoryear{{Markevitch}, {Gonzalez}, {Clowe},
  {Vikhlinin}, {Forman}, {Jones}, {Murray} \& {Tucker}}{{Markevitch}
  et~al.}{2004}]{Mar2004606}
{Markevitch} M.,  {Gonzalez} A.~H.,  {Clowe} D.,  {Vikhlinin} A.,  {Forman} W.,
   {Jones} C.,  {Murray} S.,    {Tucker} W.,  2004, \apj, 606, 819

\bibitem[\protect\citeauthoryear{{Markevitch}, {Gonzalez}, {David},
  {Vikhlinin}, {Murray}, {Forman}, {Jones} \& {Tucker}}{{Markevitch}
  et~al.}{2002}]{Mar2002567}
{Markevitch} M.,  {Gonzalez} A.~H.,  {David} L.,  {Vikhlinin} A.,  {Murray} S.,
   {Forman} W.,  {Jones} C.,    {Tucker} W.,  2002, \apjl, 567, L27

\bibitem[\protect\citeauthoryear{{Markevitch}, {Govoni}, {Brunetti} \&
  {Jerius}}{{Markevitch} et~al.}{2005}]{Mar2005627}
{Markevitch} M.,  {Govoni} F.,  {Brunetti} G.,    {Jerius} D.,  2005, \apj,
  627, 733

\bibitem[\protect\citeauthoryear{{Markevitch} \& {Vikhlinin}}{{Markevitch} \&
  {Vikhlinin}}{2001}]{Mar2001563}
{Markevitch} M.,  {Vikhlinin} A.,  2001, \apj, 563, 95

\bibitem[\protect\citeauthoryear{{Miller} \& {Owen}}{{Miller} \&
  {Owen}}{2003}]{Mil2003125}
{Miller} N.~A.,  {Owen} F.~N.,  2003, \aj, 125, 2427

\bibitem[\protect\citeauthoryear{{Mohan} \& {Rafferty}}{{Mohan} \&
  {Rafferty}}{2015}]{Moh2015}
{Mohan} N.,  {Rafferty} D., , 2015, {PyBDSM: Python Blob Detection and Source
  Measurement}, Astrophysics Source Code Library

\bibitem[\protect\citeauthoryear{{O'Dea}}{{O'Dea}}{1998}]{Ode1998110}
{O'Dea} C.~P.,  1998, \pasp, 110, 493

\bibitem[\protect\citeauthoryear{{Ogrean} \& {Br{\"u}ggen}}{{Ogrean} \&
  {Br{\"u}ggen}}{2013}]{Ogr2013433}
{Ogrean} G.~A.,  {Br{\"u}ggen} M.,  2013, \mnras, 433, 1701

\bibitem[\protect\citeauthoryear{{Owen}, {Rudnick}, {Eilek}, {Rau}, {Bhatnagar}
  \& {Kogan}}{{Owen} et~al.}{2014}]{Owe2014794}
{Owen} F.~N.,  {Rudnick} L.,  {Eilek} J.,  {Rau} U.,  {Bhatnagar} S.,
  {Kogan} L.,  2014, \apj, 794, 24

\bibitem[\protect\citeauthoryear{{Owers}, {Randall}, {Nulsen}, {Couch}, {David}
  \& {Kempner}}{{Owers} et~al.}{2011}]{Owe2011728}
{Owers} M.~S.,  {Randall} S.~W.,  {Nulsen} P.~E.~J.,  {Couch} W.~J.,  {David}
  L.~P.,    {Kempner} J.~C.,  2011, \apj, 728, 27

\bibitem[\protect\citeauthoryear{{Petrosian}}{{Petrosian}}{2001}]{Pet2001557}
{Petrosian} V.,  2001, \apj, 557, 560

\bibitem[\protect\citeauthoryear{{Planck Collaboration}, {Ade}, {Aghanim},
  {Armitage-Caplan}, {Arnaud}, {Ashdown}, {Atrio-Barandela}, {Aumont},
  {Aussel}, {Baccigalupi} \& et al.}{{Planck Collaboration}
  et~al.}{2014}]{Pla2014571}
{Planck Collaboration} {Ade} P.~A.~R.,  {Aghanim} N.,  {Armitage-Caplan} C.,
  {Arnaud} M.,  {Ashdown} M.,  {Atrio-Barandela} F.,  {Aumont} J.,  {Aussel}
  H.,  {Baccigalupi} C.,    et al. 2014, \aap, 571, A29

\bibitem[\protect\citeauthoryear{{Poggianti}, {Bridges}, {Komiyama}, {Yagi},
  {Carter}, {Mobasher}, {Okamura} \& {Kashikawa}}{{Poggianti}
  et~al.}{2004}]{Pog2004601}
{Poggianti} B.~M.,  {Bridges} T.~J.,  {Komiyama} Y.,  {Yagi} M.,  {Carter} D.,
  {Mobasher} B.,  {Okamura} S.,    {Kashikawa} N.,  2004, \apj, 601, 197

\bibitem[\protect\citeauthoryear{{Rau} \& {Cornwell}}{{Rau} \&
  {Cornwell}}{2011}]{Rau2011532}
{Rau} U.,  {Cornwell} T.~J.,  2011, \aap, 532, A71

\bibitem[\protect\citeauthoryear{{Rodriguez-Gonzalvez}, {Olamaie}, {Davies},
  {Fabian}, {Feroz}, {Franzen} \& {et al.}}{{Rodriguez-Gonzalvez}
  et~al.}{2011}]{Rod2011414}
{Rodriguez-Gonzalvez} C.,  {Olamaie} M.,  {Davies} M.~L.,  {Fabian} A.~C.,
  {Feroz} F.,  {Franzen} T.~M.~O.,    {et al.} 2011, \mnras, 414, 3751

\bibitem[\protect\citeauthoryear{{Russell}, {Fabian}, {Taylor}, {Sanders},
  {Blundell}, {Crawford}, {Johnstone} \& {Belsole}}{{Russell}
  et~al.}{2012}]{Rus2012422}
{Russell} H.~R.,  {Fabian} A.~C.,  {Taylor} G.~B.,  {Sanders} J.~S.,
  {Blundell} K.~M.,  {Crawford} C.~S.,  {Johnstone} R.~M.,    {Belsole} E.,
  2012, \mnras, 422, 590

\bibitem[\protect\citeauthoryear{{Russell}, {McNamara}, {Sanders}, {Fabian},
  {Nulsen}, {Canning}, {Baum}, {Donahue}, {Edge}, {King} \& {O'Dea}}{{Russell}
  et~al.}{2012}]{Rus2012423}
{Russell} H.~R.,  {McNamara} B.~R.,  {Sanders} J.~S.,  {Fabian} A.~C.,
  {Nulsen} P.~E.~J.,  {Canning} R.~E.~A.,  {Baum} S.~A.,  {Donahue} M.,  {Edge}
  A.~C.,  {King} L.~J.,    {O'Dea} C.~P.,  2012, \mnras, 423, 236

\bibitem[\protect\citeauthoryear{{Russell}, {Sanders}, {Fabian}, {Baum},
  {Donahue}, {Edge}, {McNamara} \& {O'Dea}}{{Russell}
  et~al.}{2010}]{Rus2010406}
{Russell} H.~R.,  {Sanders} J.~S.,  {Fabian} A.~C.,  {Baum} S.~A.,  {Donahue}
  M.,  {Edge} A.~C.,  {McNamara} B.~R.,    {O'Dea} C.~P.,  2010, \mnras, 406,
  1721

\bibitem[\protect\citeauthoryear{{Russell}, {van Weeren}, {Edge}, {McNamara},
  {Sanders}, {Fabian}, {Baum}, {Canning}, {Donahue} \& {O'Dea}}{{Russell}
  et~al.}{2011}]{Rus2011417}
{Russell} H.~R.,  {van Weeren} R.~J.,  {Edge} A.~C.,  {McNamara} B.~R.,
  {Sanders} J.~S.,  {Fabian} A.~C.,  {Baum} S.~A.,  {Canning} R.~E.~A.,
  {Donahue} M.,    {O'Dea} C.~P.,  2011, \mnras, 417, L1

\bibitem[\protect\citeauthoryear{{Shimwell}, {Brown}, {Feain}, {Feretti},
  {Gaensler} \& {Lage}}{{Shimwell} et~al.}{2014}]{Shi2014440}
{Shimwell} T.~W.,  {Brown} S.,  {Feain} I.~J.,  {Feretti} L.,  {Gaensler}
  B.~M.,    {Lage} C.,  2014, \mnras, 440, 2901

\bibitem[\protect\citeauthoryear{{Taylor}, {Ellingson}, {Kassim}, {Craig},
  {Dowell} \& {et al.}}{{Taylor} et~al.}{2012}]{Tay2012}
{Taylor} G.~B.,  {Ellingson} S.~W.,  {Kassim} N.~E.,  {Craig} J.,  {Dowell} J.,
     {et al.} 2012, Journal of Astronomical Instrumentation, 1, 1250004

\bibitem[\protect\citeauthoryear{{van Weeren}, {Andrade-Santos}, {Dawson},
  {Golovich}, {Lal} \& {et al.}}{{van Weeren} et~al.}{2017}]{vanW2017Nat}
{van Weeren} R.~J.,  {Andrade-Santos} F.,  {Dawson} W.~A.,  {Golovich} N.,
  {Lal} D.~V.,    {et al.} 2017, Nature Astronomy, 1, 0005

\bibitem[\protect\citeauthoryear{{van Weeren}, {Br{\"u}ggen}, {R{\"o}ttgering}
  \& {Hoeft}}{{van Weeren} et~al.}{2011}]{vanW2011418}
{van Weeren} R.~J.,  {Br{\"u}ggen} M.,  {R{\"o}ttgering} H.~J.~A.,    {Hoeft}
  M.,  2011, \mnras, 418, 230

\bibitem[\protect\citeauthoryear{{van Weeren}, {Brunetti}, {Br{\"u}ggen},
  {Andrade-Santos} \& {et al.}}{{van Weeren} et~al.}{2016}]{vanW2016818}
{van Weeren} R.~J.,  {Brunetti} G.,  {Br{\"u}ggen} M.,  {Andrade-Santos} F.,
  {et al.} 2016, \apj, 818, 204

\bibitem[\protect\citeauthoryear{{van Weeren}, {Intema}, {Lal},
  {Andrade-Santos}, {Br{\"u}ggen}, {de Gasperin}, {Forman}, {Hoeft}, {Jones},
  {Nuza}, {R{\"o}ttgering} \& {Stroe}}{{van Weeren} et~al.}{2014}]{vanW2014786}
{van Weeren} R.~J.,  {Intema} H.~T.,  {Lal} D.~V.,  {Andrade-Santos} F.,
  {Br{\"u}ggen} M.,  {de Gasperin} F.,  {Forman} W.~R.,  {Hoeft} M.,  {Jones}
  C.,  {Nuza} S.~E.,  {R{\"o}ttgering} H.~J.~A.,    {Stroe} A.,  2014, \apjl,
  786, L17

\bibitem[\protect\citeauthoryear{{van Weeren}, {R{\"o}ttgering}, {Br{\"u}ggen}
  \& {Hoeft}}{{van Weeren} et~al.}{2010}]{vanW2010Sci}
{van Weeren} R.~J.,  {R{\"o}ttgering} H.~J.~A.,  {Br{\"u}ggen} M.,    {Hoeft}
  M.,  2010, Science, 330, 347

\bibitem[\protect\citeauthoryear{{Venturi}, {Giacintucci}, {Dallacasa},
  {Cassano}, {Brunetti}, {Macario} \& {Athreya}}{{Venturi}
  et~al.}{2013}]{Ven2013551}
{Venturi} T.,  {Giacintucci} S.,  {Dallacasa} D.,  {Cassano} R.,  {Brunetti}
  G.,  {Macario} G.,    {Athreya} R.,  2013, \aap, 551, A24

\bibitem[\protect\citeauthoryear{{White}, {Canning}, {King}, {Lee}, {Russell},
  {Baum}, {Clowe}, {Coleman}, {Donahue}, {Edge}, {Fabian}, {Johnstone},
  {McNamara}, {O'Dea} \& {Sanders}}{{White} et~al.}{2015}]{Whi2015453}
{White} J.~A.,  {Canning} R.~E.~A.,  {King} L.~J.,  {Lee} B.~E.,  {Russell}
  H.~R.,  {Baum} S.~A.,  {Clowe} D.~I.,  {Coleman} J.~E.,  {Donahue} M.,
  {Edge} A.~C.,  {Fabian} A.~C.,  {Johnstone} R.~M.,  {McNamara} B.~R.,
  {O'Dea} C.~P.,    {Sanders} J.~S.,  2015, \mnras, 453, 2718

\bibitem[\protect\citeauthoryear{{Zwart}, {Barker}, {Biddulph}, {Bly},
  {Boysen}, {Brown}, {Clementson}, {Crofts} \& {et al.}}{{Zwart}
  et~al.}{2008}]{Zwa2008391}
{Zwart} J.~T.~L.,  {Barker} R.~W.,  {Biddulph} P.,  {Bly} D.,  {Boysen} R.~C.,
  {Brown} A.~R.,  {Clementson} C.,  {Crofts} M.,    {et al.} 2008, \mnras, 391,
  1545

\end{thebibliography}
